\documentclass[twocolumn,secnumarabic,amssymb, nobibnotes, aps, pra]{revtex4-2}

\usepackage{graphicx}
\usepackage{amsmath}
\usepackage{xcolor}
\usepackage[dvipsnames]{xcolor}
\usepackage[normalem]{ulem}

\usepackage{multirow}

\usepackage{booktabs}
\begin{document}
	
	\title{Electric Field Distortions in Surface Ion Traps with Integrated Nanophotonics}%
	
	\author{Guochun Du$^{1}$}
		\email[Contact author: ]{guochun.du@ptb.de}
		 	\author{Elena Jordan$^{1}$,  Tanja E. Mehlst\"aubler$^{1,2,3}$}%

	\affiliation{$^1$ Physikalisch-Technische Bundesanstalt, Bundesallee 100, 38116 Braunschweig, Germany\\ 
		$^2$ Institut f\"ur Quantenoptik, Leibniz Universit\"at Hannover, Welfengarten 1, 30167 Hannover, Germany\\
		$^3$ Laboratory of Nano and Quantum Engineering, Leibniz Universtit\"at Hannover, Schneiderberg 39, 30167 Hannover, Germany}
	\date{October 2025}%
	
	\begin{abstract}

		The integration of photonic components into surface ion traps provides a scalable approach for trapped-ion quantum computing, sensing, and metrology, enabling compact systems with enhanced stability and precision. However, the introduction of optical apertures in the trap electrodes can distort the trapping electric field. This can lead to excess micromotion (EMM) and ion displacement which degrade the performance of quantum logic operations and optical clocks. In this work, we systematically investigate the electric field distortion in a surface ion trap with integrated waveguides and grating couplers that require apertures in the surface electrode using Finite Element Method (FEM) simulations. We analyze methods to reduce these distortions by exploiting symmetries and  transparent conductive oxide materials.
		
	\end{abstract}
	
	\maketitle
	
	\section{Introduction}
	
	Ion traps are a key technology for trapped-ion based quantum computers, quantum simulators, quantum sensors, and quantum metrology \cite{Bruz2019, Blatt2012, Dege2017, Giov2004}.  Currently, the number of ions for these applications is limited by the trap technology and the laser optics that is needed for laser cooling, state control, and spectroscopy \cite{Mose2023, Pogo2021}. Integrating optics monolithically into ion traps is a viable way to make ion traps scalable to store larger numbers of qubits, to reduce the size of the experimental setup, and at the same time  enhance the pointing stability \cite{Zesa2024, sotirova2024,Day2021, ghadimi2017,Mehta2016,Mehta2020,Mordi2025, Niff2020, Ivory2021, Kwon2024}. The stability and compactness make them particularly suitable for portable quantum sensors. 
	
	Recent advances in the fabrication of nanophotonics, especially for the blue and UV wavelength range \cite{West2019, Hend2023}, pave the way for the integration of all wavelengths needed to control many ion species. Optically integrated ion traps with waveguides and grating outcouplers \cite{Beck2024} to deliver light to the ions have been demonstrated for single-ion addressing \cite{Mehta2016},  excitation of optical clock transitions \cite{Mehta2016,Niff2020, Ivory2021, Kwon2024}, and multi-ion quantum logic operations \cite{Mehta2020, Mordi2025}. For some demonstrators increased heating rates have been measured \cite{Mehta2020, Niff2020, Mordi2025}.

	Previous simulation-based studies have shown how integrated micro-optics and miniature optical cavities can perturb trapping fields in ion traps \cite{jordan2025scalable, podoliak2016comparative, kassa2025integrate}. Motivated by these works, we systematically investigate, using Finite Element Method (FEM) simulations, how the integration of nanophotonics with grating couplers distorts the trapping electric field of a surface ion trap.  If the amplitude of the trapping RF field cannot be zeroed at the position of the ion due to this distortion,  the residual RF field amplitude leads to excess micromotion (EMM). In an anharmonic trapping potential, EMM causes heating of the ion's secular motion \cite{Wine1997, Berkeland1998}. In a harmonic trapping potential,  the gradient of a residual RF field may result in increased heating rates \cite{Kali2021, Blakestad2009} and heating during shuttling operations \cite{Mordi2025}. EMM also contributes to frequency shifts, such as  time dilation shift  and AC Stark shift,  which often limit the accuracy of trapped ion based optical clocks \cite{Kell2019}. Finally, the distortion can lead to ion displacement from the target position, causing the outcoupled beam to be misaligned with the ion. 
	
	In Section 2, we introduce the electrode geometry of the surface ion trap with apertures for grating couplers and the simulation setup for analyzing the RF field distortion. Section 3 examines the design constraints imposed by the grating couplers.  In Section 4, the simulation results of the RF field distortion caused by the aperture are presented. Approaches to mitigate the distortion through the application of symmetry principles and the use of transparent conductive oxide materials are then examined. In Section \ref{Effect of varying the conductivity of TCO}  and \ref{Effect of the RF phase shift}, the effect of a possible phase shift of the RF field is also discussed. Section 5 concludes by summarizing the effects of the aperture, offering design suggestions, and outlining future perspectives to mitigate distortion.
	
	\section{Electrode Geometry and Simulation Setup}
	\label{Electrode Geometry and Simulation Setup}
	
\begin{figure}[b]
		\centering
		\includegraphics[scale=0.094]{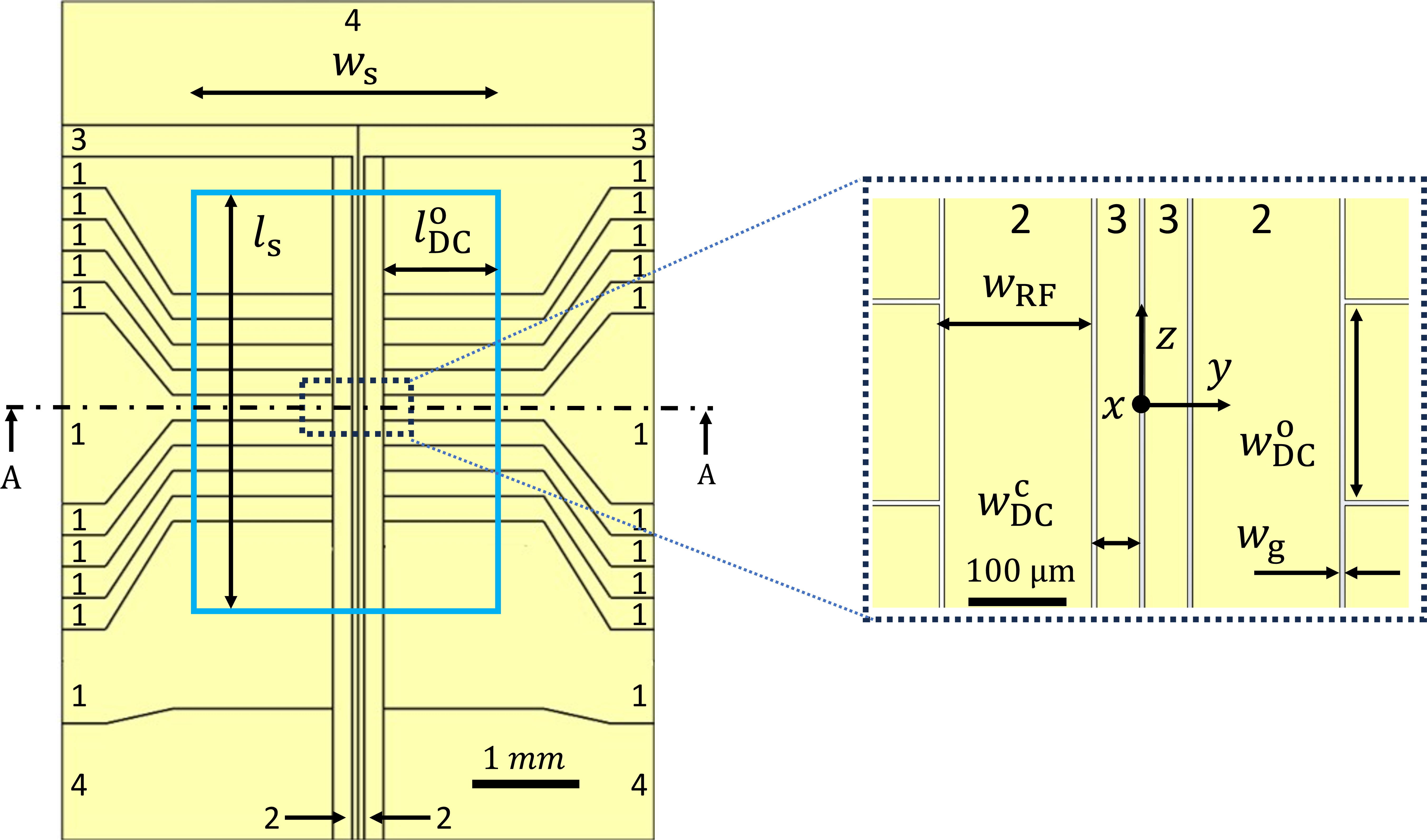}
		\caption{Top view of the surface trap studied in the simulations. Label 1 identifies the outer DC electrodes, label 2 the RF electrodes, label 3 the center DC electrodes and label 4 the ground electrodes. Only the region within the blue rectangle is considered in the simulation. The origin of the coordinate system (referred to as the trap center in the following discussion) is positioned at the center of the simulation region in the yz-plane, while in x-direction, it is located on the surface of the gold electrode.  Unless otherwise stated, all simulations shown in this paper use the same coordinate system as shown here. The trapping field is generated by applying a RF potential $U_\mathrm{RF}$ to the RF electrodes (2) and DC potentials $U_\mathrm{DC}$ to the DC electrodes (1, 3). The Symbol A indicates a cutting plane for a cross-sectional view shown in figure \ref{fig:cross_view_ion_trap_aperture}(a).}
		\label{fig:top_view_ion_trap}
		
	\end{figure}

	For the discussions in this paper, we use a design of a surface ion trap 
	from reference \cite{Allc2010}.   We modified the width of the center DC  electrodes and RF  electrodes to confine Yb$^{+}$ ions at a height of 100 $\mu$m above the surface on the trap axis which is defined as the central axis of the trap in z-direction at the target ion height. Figure \ref{fig:top_view_ion_trap} shows a top view of the electrode geometry. \\

	\begin{table}[h]
		\centering
		\caption{Dimensions of the surface trap geometry and simulation setup.}
		\footnotesize
		\begin{tabular}{ccc}
			\hline
			Symbol & Value & Description \\
			\hline
			$l_\mathrm{s}$ & 5 mm & length of the simulation region \\
			$w_\mathrm{s}$ & 2.4 mm & width of the simulation region \\
			$l^\mathrm{o}_\mathrm{DC}$ & 1 mm & length of the outer DC electrode \\
			$w^\mathrm{o}_\mathrm{DC}$ & 200 $\mu$m & width of the outer DC electrode \\    
			$w_\mathrm{RF}$ & 150 $\mu$m & width of the RF electrode \\
			$w^\mathrm{c}_\mathrm{DC}$ & 44.3 $\mu$m & width of the center DC electrode \\
			$w_\mathrm{g}$ & 5 $\mu$m & width of the gap between electrodes \\
			$t_\mathrm{e}$ & 6 $\mu$m & thickness of the gold electrode layer \\
			$t_\mathrm{d}$ & 10 $\mu$m & thickness of the dielectric (SiO$_2$) layer \\
			$t_\mathrm{gp}$ & 3 $\mu$m & thickness of the ground plane \\
			$t_\mathrm{c}$ & 7 $\mu$m & thickness of the SiO$_2$ cladding \\
			$t_\mathrm{w}$ & 525 $\mu$m & thickness of the wafer \\
			\hline
		\end{tabular}
		\label{tab:dimensions}
	\end{table}

	The height of 100 µm is selected to balance the trade-off between mitigating surface-induced heating \cite{Epstein2007,Deslauriers2006} and maintaining sufficient trapping depth for stable ion confinement. The trapping field is generated by the application of a RF (radio frequency) potential and DC (direct current) potentials to the corresponding electrodes. Given a RF amplitude of $U_\mathrm{RF} = 100 $ V and a RF frequency of $\frac{\Omega_\mathrm{rf}}{2\pi} = 16$ MHz, we obtain a secular frequency of $\nu_\mathrm{sec} = 1.9 $ MHz and a trapping depth of 103.01 meV for a $^{172}$Yb$^+$ ion. The relevant dimensions of the electrodes are detailed in Table \ref{tab:dimensions}.

	\begin{figure}[t]
		
		\centering
		
		\includegraphics[scale=0.06]{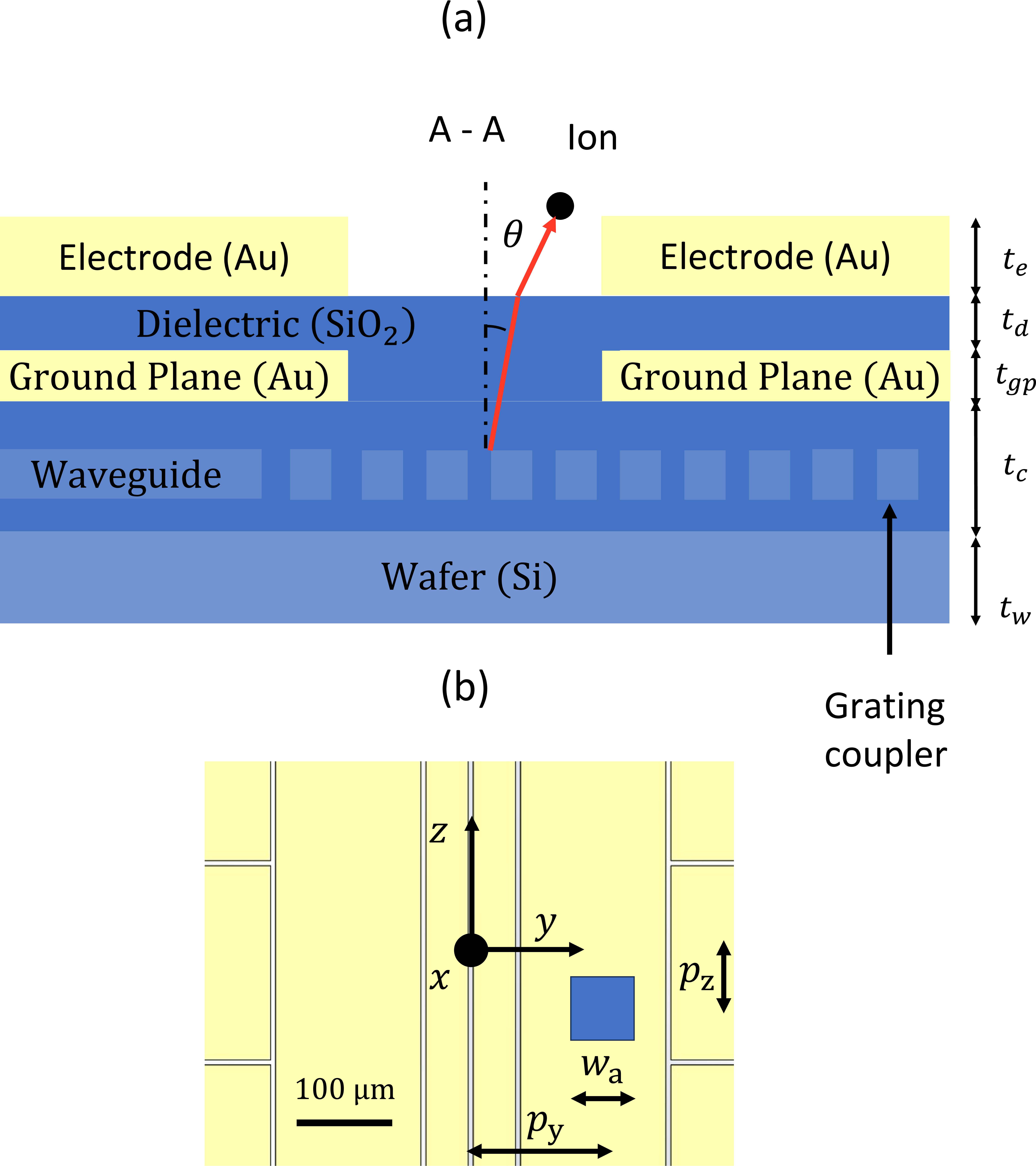}
		
		\caption{(a) A cross-sectional view (A-A) of the surface trap reveals the layer stack-up of the photonic integrated ion trap chip. The cross-sectional view is schematic and not drawn to scale and the layer thicknesses are given in Table \ref{tab:dimensions}.
			A ground plane positioned  3 $\mathrm{\mu}$m beneath the trap electrodes shields the silicon substrate from the RF trapping fields. The aperture in the ground plane has the same width as in the electrode. The waveguide layer is embedded within the SiO$_\mathrm{2}$ cladding layer on a silicon wafer. All the layers are taken into account in the simulations shown in this paper except the waveguide layer, as its contribution to the field distribution at the ion's position is negligible.  The angle of the outcoupled laser beam relative to the surface normal (x-axis) is denoted by $\theta$. (b) Top view of an square-shaped aperture with a size of  $w_\mathrm{a}\times w_\mathrm{a}$ in the gold electrode. $p_\mathrm{z}$ indicates the position of the center of the aperture, while $p_\mathrm{y}$ indicates the position in y-direction.}
		\label{fig:cross_view_ion_trap_aperture}
	\end{figure}

	To integrate optical components into the surface trap, we adapted the layer stack-up from  reference \cite{Mehta2020} in the design as depicted in figure \ref{fig:cross_view_ion_trap_aperture}. A waveguide layer with grating outcouplers is implemented beneath the electrodes. Light is coupled into this layer, guided, and subsequently coupled out vertically by grating outcouplers through apertures in the gold electrodes. All apertures shown in the schematic figures throughout this paper are drawn for illustrative purposes only and are not to scale, as both their size and position are treated as variable parameters in the simulations.

	The discontinuity arising from the aperture can introduce a distortion of the trapping field. 
	We investigate the distortion of the trapping RF field using FEM simulations with the commercial software \textit{COMSOL Multiphysics 5.6}.  All simulations utilize  square-shaped apertures. 
	The target ion position for a single ion is 100 $\mathrm{\mu}$m directly above the trap center.  
	The ion's position in the radial direction (in  x-y plane), which is perpendicular to the trap axis, is determined by the minimum of the radial RF field $E_\mathrm{rf,r}$ in this plane, where $E_\mathrm{rf,r} = \sqrt{{E_\mathrm{rf,x}}^2 + {E_\mathrm{rf,y}}^2}$. 
	This position is influenced by the electrode geometry and is independent on the magnitude of the RF field. 	Due to the finite length of the trap, a residual RF field exists along the trap axis. This residual field scales with the magnitude of the applied RF field. Accordingly, in the following discussion, we consider distortions in the radial and axial directions separately. In the radial direction, we analyze the displacement of the $E_\mathrm{rf,r}$ minimum from the target ion position, while in the axial direction, we examine the residual RF field along the trap axis. To simplify the simulation of the trapping  field, it is initially sufficient to consider the RF field at a fixed phase, which reduces the problem to an electrostatic analysis \cite{Hers2012}.  For the simulations presented in the following sections,  we set $U_\mathrm{RF} = 100 $ V and keep all $U_\mathrm{DC}  = 0$ V unless otherwise stated.  The ground plane is modeled as an equipotential surface held at 0~V. The silicon substrate is treated as a dielectric material, consistent with a high-resistivity Si substrate and the absence of a backside ground plane in the modeled geometry. The relative permittivities used in the simulations are $\varepsilon_r = 3.9$ for SiO$_2$ and $\varepsilon_r = 11.7$ for Si.

	\section{Design constraints imposed by grating couplers}
	
	The grating couplers constrain the position and the geometry of the apertures in the trap electrodes. The light from the waveguide is coupled out at an angle $\theta$ by a grating coupler when the Bragg condition,

	\begin{equation}
		n_\mathrm{eff}  -  \mathrm{sin}\theta= \frac{m\lambda}{\Lambda},
	\end{equation}
	
	\noindent{is satisfied, where  $m$ is the diffraction order, $\lambda$ is the wavelength, $\Lambda$ is the period of the grating and $n_\mathrm{eff}$ is the effective refractive index of the grating.} The effective refractive index can be approximated as a weighted average of the refractive indices associated with the guided mode in the waveguide, \( n_\mathrm{WG} \), and the cladding region, \( n_\mathrm{clad} \), according to
	\begin{equation}
		n_{\mathrm{eff}}^{2} = f\,n_\mathrm{WG}^{2} + (1-f)\,n_\mathrm{clad}^{2}.
	\end{equation}
	Here, the duty cycle \( f \) represents the fraction of the waveguide region within one grating period and is defined as \( f = l_p / \Lambda \), where \( l_p \) is the length of the waveguide region and \( \Lambda \) is the grating period.

\begin{figure}[!htbp]
		
		\centering
		
		\includegraphics[scale=0.5]{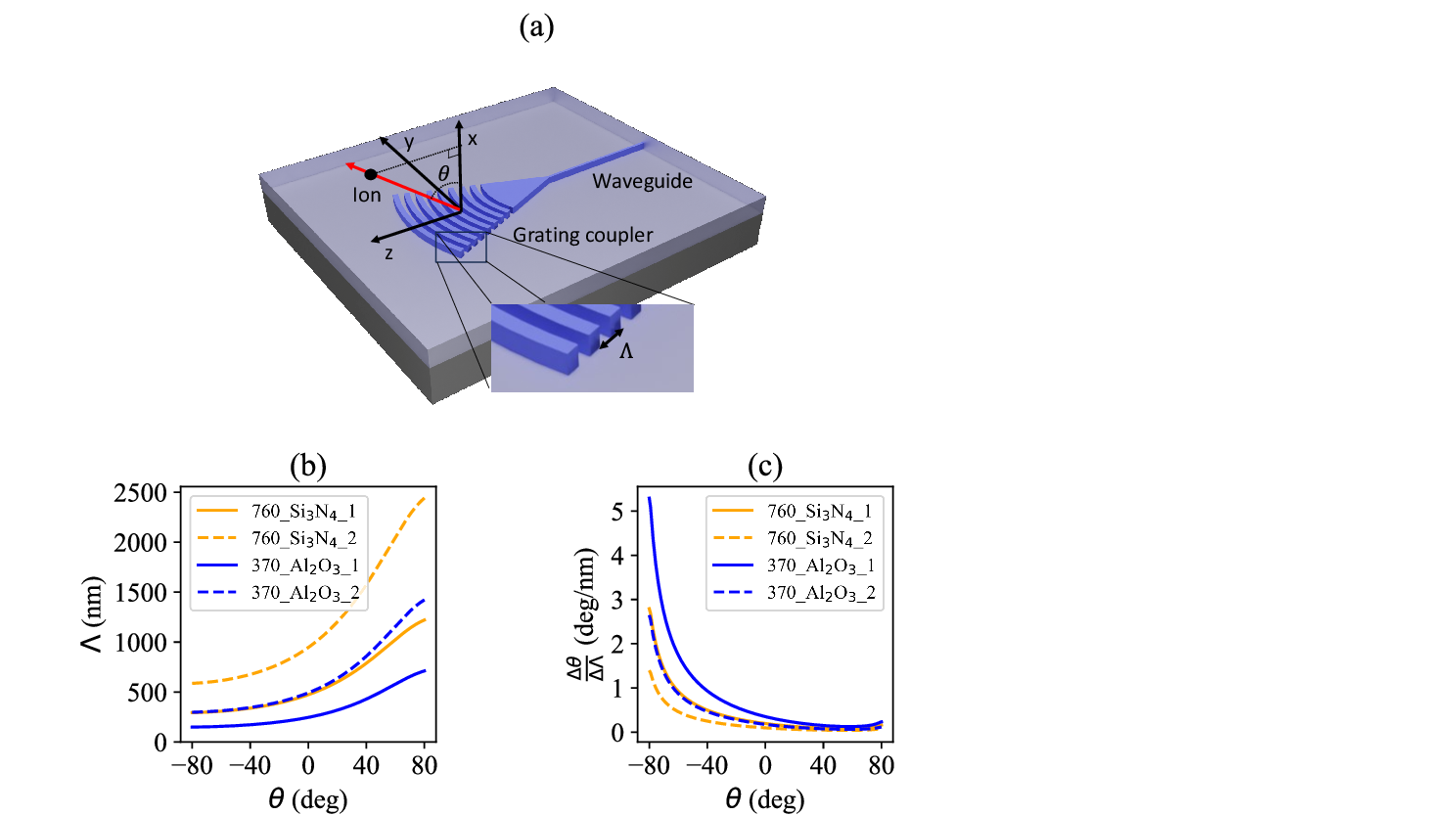}
		
		\caption{(a) A grating coupler with period $\Lambda$ couples light out from a waveguide at an angle $\theta$ onto an ion in free space. Here a forward grating coupler with positive outcoupling angle $\theta$ is displayed. (b) The grating period $\Lambda$ and (c) the angle-to-period sensitivity $\frac{\Delta\theta}{\Delta\Lambda}$ for the first- and second-order beams as functions of $\theta$. The first-order beams are represented by solid lines, while the second-order beams are shown with dashed lines. Data are shown for a Si$_3$N$_4$ grating with a wavelength of 760 nm and a thickness of 200 nm, and an Al$_2$O$_3$ grating with a wavelength of 370 nm and a thickness of 120 nm.}
		\label{fig:grating_coupler}
	\end{figure}

	Figure \ref{fig:grating_coupler}(a) illustrates the structure of a grating coupler. A backward grating coupler emits in the direction opposite to the propagation of the guided light, resulting in a negative diffraction angle $\theta$, while a forward grating coupler diffracts the guided light with a positive diffraction angle $\theta$. As illustrated in figures \ref{fig:grating_coupler}(b) and (c), the grating period decreases as $\theta$ transitions from positive to negative values,  while the absolute angle-to-period sensitivity $\frac{\Delta\theta}{\Delta\Lambda}$ increases. Additionally, shorter wavelengths require smaller grating periods and exhibit higher angle-to-period sensitivity. Another consideration for trap design is that, backward grating couplers emit only first-order diffraction, while forward grating couplers can generate both first and higher-order diffracted beams. 
	
	For example, for a forward Si$_3$N$_4$ grating coupler with a thickness of $200~\mathrm{nm}$ at a wavelength of $760~\mathrm{nm}$, a first-order diffraction angle of $\theta = 20^\circ$ corresponds to a grating period of approximately $600~\mathrm{nm}$. At this period, the second-order diffraction condition is also satisfied, producing an additional beam at a backward angle of about $-67^\circ$. As the first-order diffraction angle $\theta$ increases, which corresponds to a further increase of the grating period, the emission angle of this second-order beam shifts continuously toward zero and can eventually become forward-directed. These higher order beams can cause unwanted stray light in the ion trap.  As a result, these considerations ultimately constrain the placement of apertures through the minimum grating feature size, the tolerance on the outcoupling angle, and the potential impact of stray light.

	\section{Results}
	
	Field distortions arising from apertures in the electrode in the context of surface-electrode ion traps with integrated gratings couplers  have been discussed previously in \cite{mehta2019towards}. Here we systematically investigate the dependence of these distortions on aperture type, position, and geometry, as well as strategies for their mitigation.
	In this section, we first discuss the RF field distortion resulting from a single aperture in different types of trap electrodes. Next, we examine how the position and geometry of an aperture, specifically in a RF electrode, affect this distortion. Following this, we investigate methods to reduce the distortion by utilizing symmetry and applying transparent conductive oxide materials. 
	Throughout the discussion, the configuration without any aperture in the trap electrodes is treated as the reference configuration, providing a baseline to evaluate the effects of apertures on the RF field distortion.

	\subsection{Effect of a single aperture  in the RF electrode, the center DC electrode, or the outer DC electrode}
	
\begin{figure}[!htbp]
		
		\centering
		\includegraphics[scale=0.043]{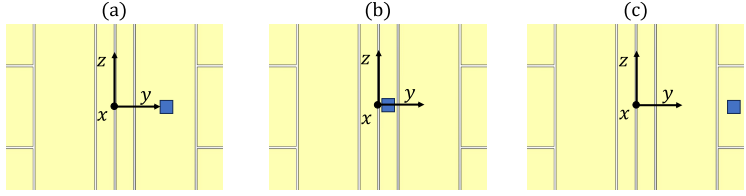}
		\caption{Layout of an aperture with a size of $w_\mathrm{a}$ = 30 $\mathrm{\mu}$m positioned in: (a) the middle of RF electrode with $p_\mathrm{z}$ = 0 and $p_\mathrm{y}$ = 126.8 $\mathrm{\mu}$m, (b) the middle of the center DC electrode with $p_\mathrm{z}$ = 0 and $p_\mathrm{y}$ = 24.7 $\mathrm{\mu}$m and (c) the outer DC electrode with $p_\mathrm{z}$ = 0 and $p_\mathrm{y}$ = 257 $\mathrm{\mu}$m. These positions correspond to outcoupling angles $\theta$ of $52^\circ$, $14^\circ$, and $69^\circ$.}
		\label{fig:single_hole_electrode_configuration}
		
	\end{figure}

	We begin by investigating the electric field distortion caused by a square-shaped aperture with width $w_\mathrm{a}$ = 30 $\mathrm{\mu}$m positioned in either the RF electrode, center DC electrode, or outer DC electrode (figure \ref{fig:single_hole_electrode_configuration}). These positions correspond to outcoupling angles $\theta$ ranging from $14^\circ$ to $69^\circ$, encompassing the range of angles demonstrated in previously reported ion traps with integrated waveguides and grating couplers \cite{Mehta2016,Niff2020, Ivory2021, Kwon2024, Mehta2020, Mordi2025}.\\\\
	
	\noindent{\textit{Shift of the RF field minimum in radial direction}}

	\noindent{The unperturbed radial RF electric field $E_\mathrm{rf,r}$ ($E_\mathrm{rf,r}  = \sqrt{{E_\mathrm{rf,x}}^2+{E_\mathrm{rf,y}}^2}$) in  x-y plane for the reference configuration is illustrated in figure \ref{fig:compare_minimum_Erf_single_hole}(a). When an aperture  
		
		\begin{figure}[!htbp]

			\centering
			\includegraphics[scale=0.2]{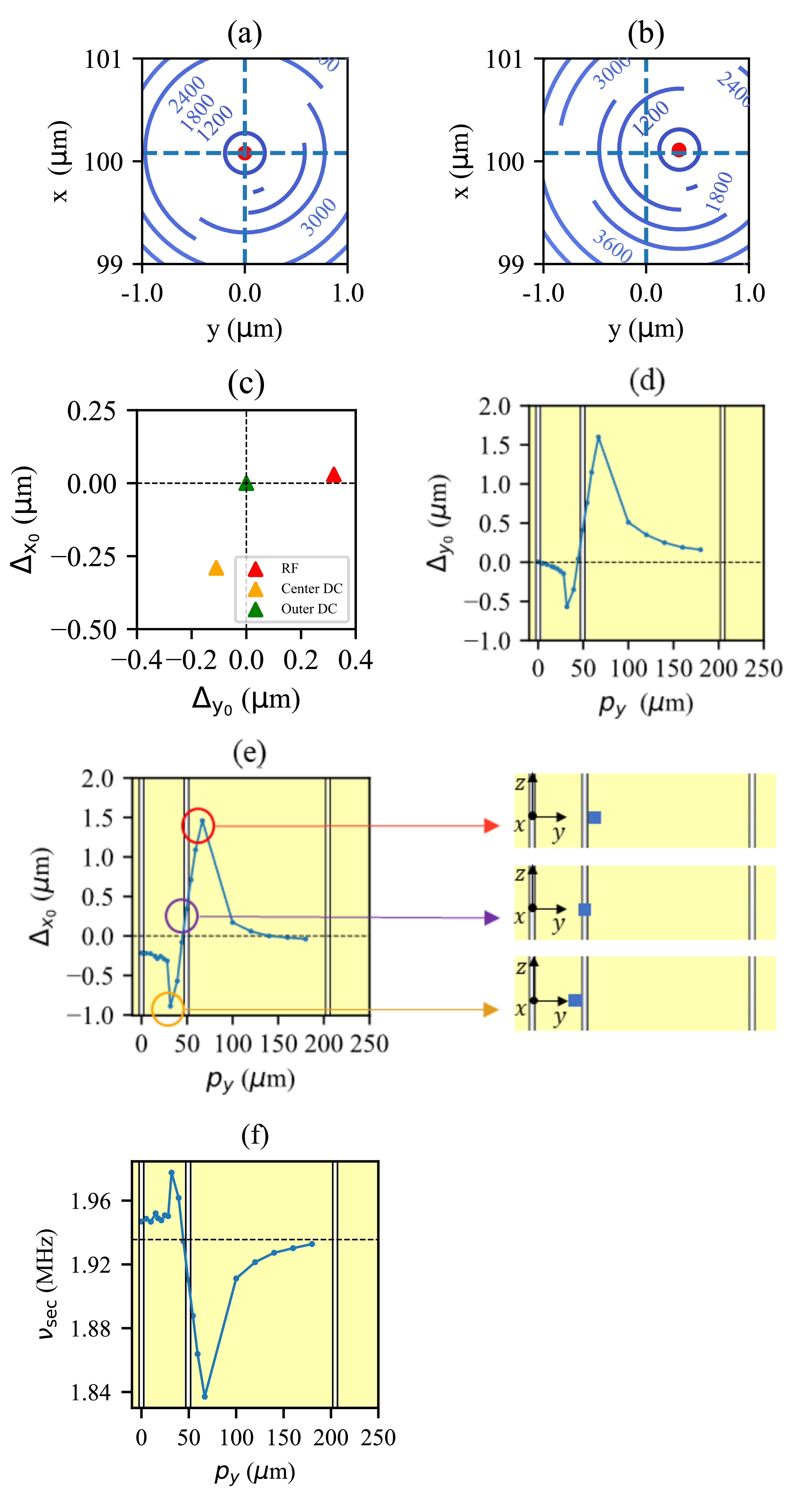}
			\caption{(a) Radial RF field $E_\mathrm{rf,r}$   in x-y plane. The values in the contour plot represent the amplitude of $E_\mathrm{rf,r}$ in V/m, the red dot the field minimum. (b) Radial RF field $E_\mathrm{rf,r}$ in V/m in x-y plane with aperture in the RF electrode, see figure \ref{fig:single_hole_electrode_configuration}(a). (c) Displacement of  $E_\mathrm{rf,r}$  minimum for apertures located in the RF electrode, the center DC electrode and the outer DC electrode as shown in figure \ref{fig:single_hole_electrode_configuration}(a), (b) and (c). Subplots (d) and (e) show the displacement of the $E_\mathrm{rf,r}$  minimum in x- and y-direction while moving an aperture in y-direction from the center DC electrode to the RF electrode and keeping $w_\mathrm{a}$ = 30 $\mathrm{\mu}$m and $p_\mathrm{z}$ = 0 $\mathrm{\mu}$m. Subplot (f) shows the corresponding secular frequency $\nu_\mathrm{sec}$ evaluated at the $E_\mathrm{rf,r}$ minimum for a $^{172}\mathrm{Yb}^+$ ion. The black dashed line denotes the reference configuration without an aperture.}
			\label{fig:compare_minimum_Erf_single_hole}
		\end{figure}

		\noindent{with a size  of $w_\mathrm{a}$ = 30 $\mathrm{\mu}$m is present in the RF electrode, see figure \ref{fig:single_hole_electrode_configuration}(a), the radial electric field $\mathrm{E_\mathrm{rf,r}}$ is shifted in  x-y plane as illustrated in figure \ref{fig:compare_minimum_Erf_single_hole}(b). The radial field $E_\mathrm{rf,r}$ minimum is displaced by  320 nm from the target ion position in y-direction and 30 nm in x-direction.}

		Figure \ref{fig:compare_minimum_Erf_single_hole}(c) provides a summary of the $E_\mathrm{rf,r}$ minimum displacements in both x and y-directions for an aperture in the RF electrode, the center DC electrode, and the outer DC electrode. 	When the aperture is in the outer DC electrode, the $E_\mathrm{rf,r}$ minimum experiences a less significant displacement compared to the RF or the center DC electrode. 
		When an aperture is located in the RF electrode, the $E_\mathrm{rf,r}$ minimum mainly shifts in positive y-direction. In contrast, when the aperture is placed in the center DC electrode, the displacement reverses its direction, and the shift in x-direction becomes more prominent.

		We resolve the transition of the displacement as the aperture is moved from the RF electrode to the center DC electrode, depicted in figure \ref{fig:compare_minimum_Erf_single_hole} (d-e). The displacement reaches its absolute maximum when the aperture is located such that one side of the aperture coincides with the edge of the electrode. A zero displacement occurs when the aperture is positioned in the gap between electrodes due to a sign change in the displacement. As the aperture moves closer to the trap center, the displacement magnitude decreases. When the aperture is directly beneath the target ion position, the  displacement magnitude in x-direction reaches a minimum of 200 nm, while in y-direction it vanishes due to the trap's symmetry with respect to the z-axis.
		
		In addition to the displacement of the RF field minimum, the presence of the aperture also modifies the curvature of the effective RF pseudopotential. We evaluate the radial secular frequency at the position of the $E_\mathrm{rf,r}$ minimum, see figure \ref{fig:compare_minimum_Erf_single_hole}(f). The secular frequency increases when the aperture is located in the center DC electrode and decreases when it is located in the RF electrode. The magnitude of the frequency change follows the same trend as the displacement of the RF field minimum, reaching a maximum relative change of approximately 5\% when the displacement is largest, i.e., when one side of the aperture coincides with the edge of the electrode.
		\\

		\noindent{\textit{Residual RF field on the trap axis}}


		\noindent In the axial direction we evaluate the RF field along the trap axis, as shown in figure ~\ref{fig:compare_all_axial}. 
		For the reference configuration without apertures, the residual RF field arising from the finite length of the trap is weak and varies smoothly along the trap axis (black dashed line). 
		Due to the symmetry of the ion trap with respect to the z-axis, the y-component vanishes along the entire trap axis. 
		The x-component exhibits a weak parabolic dependence, while the z-component varies linearly and remains zero at the trap center ($z=0$), with gradients on the order of several tens of $\mathrm{mV/mm^2}$.
		
		When an aperture is introduced in the RF electrode (see red curve in figure \ref{fig:compare_all_axial}),  the residual RF field along the trap axis is strongly enhanced, with the dominant contributions appearing in the y- and z-components. 
		The y-component shows a pronounced peak at the trap center. In this case, the y-component reaches a peak amplitude of $994\,\mathrm{V/m}$ at $z = 0$, with a gradient of $5.5\,\mathrm{V/mm^2}$ at the FWHM, while the z-component has a gradient of approximately $4.3\,\mathrm{V/mm^2}$ at $z = 0$.
		
		In contrast, when the aperture is located in the center DC electrode, the residual RF field is redistributed, and the dominant distortions occur in the x- and z-components. 
		The x-component shows a pronounced peak at the trap center, while the z-component again displays an anti-symmetric dispersive lineshape. 
		The resulting peak field amplitudes and axial gradients are of similar order to those observed for the y- and z-components when the aperture is placed in the RF electrode.

		\begin{figure}[t]
			\centering
			\includegraphics[scale=0.48]{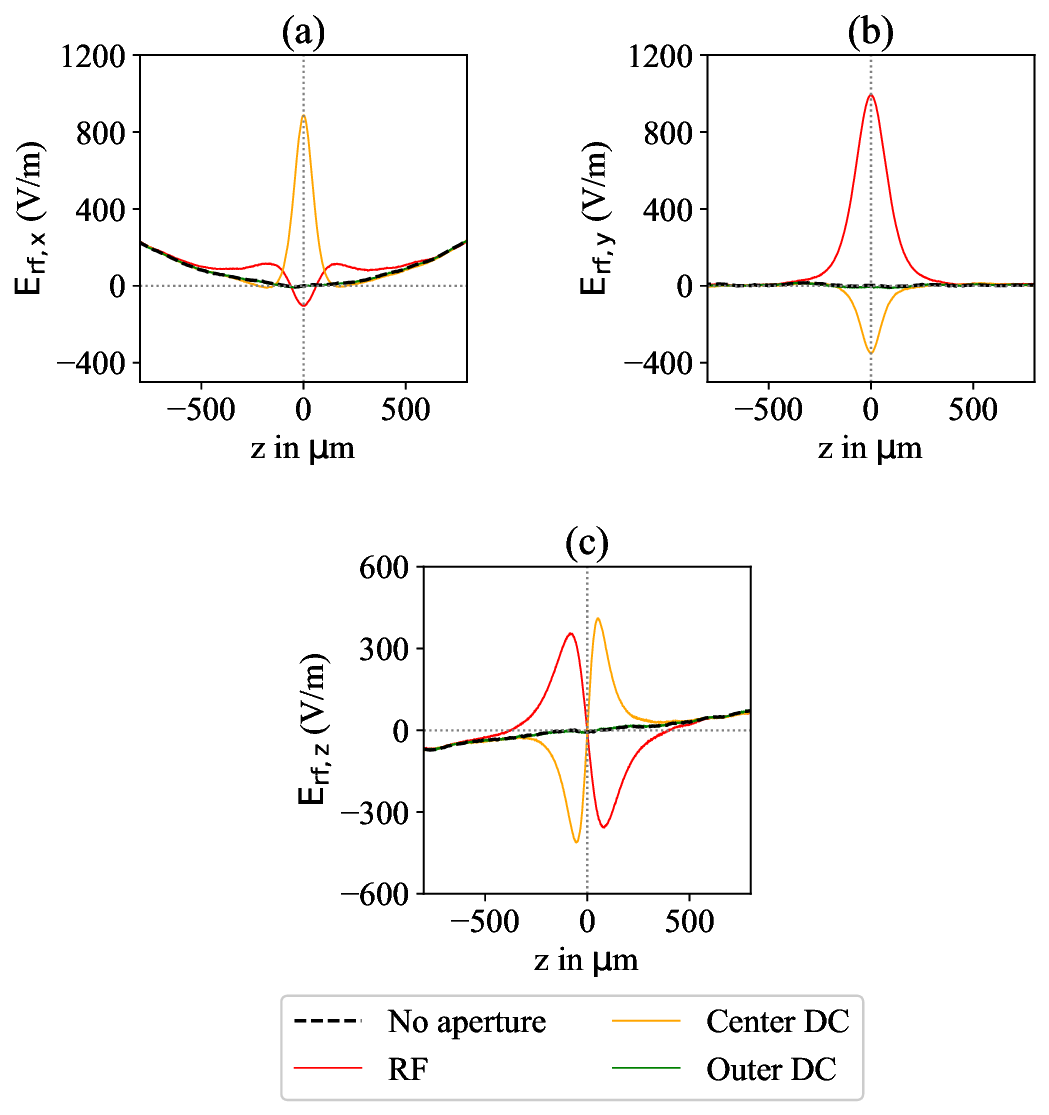}
			\caption{Subplots (a)-(c) show the x-, y- and z-components of the RF field along the trap axis  for different electrode configurations: the black dashed line represents the reference configuration, the red line represents a single  aperture in the RF electrode as shown in figure \ref{fig:single_hole_electrode_configuration}(a), the orange line represents the aperture in the center DC electrode as shown in figure \ref{fig:single_hole_electrode_configuration}(b)  and the green line represents the aperture in the outer DC electrode as shown in figure \ref{fig:single_hole_electrode_configuration}(c). The black and green lines overlap which demonstrates the small impact of an aperture in the outer DC electrode. 
			}
			\label{fig:compare_all_axial}
		\end{figure}

		The differences in the radial components $E_\mathrm{rf,x}$ and $E_\mathrm{rf,y}$ at the target ion position (z = 0) explain the behavior of the displacement of the $E_\mathrm{rf,r}$ shown in figure \ref{fig:compare_minimum_Erf_single_hole}(c). At the target ion position,  for an aperture in the RF electrode $E_\mathrm{rf,y}$ dominates over $E_\mathrm{rf,x}$, while for an aperture in the center DC electrode, $E_\mathrm{rf,x}$ becomes dominant. Additionally, both $E_\mathrm{rf,x}$ and $E_\mathrm{rf,y}$ switch  signs depending on the aperture's location. This explains why the $E_\mathrm{rf,r}$ minimum shifts positively, with the displacement more prominent in y-direction, when the aperture is in the RF electrode. For the case  the aperture is in the center DC electrode, the displacement reverses direction, with x-direction shift becoming more pronounced.

		To summarise the comparison of the three electrode types,  placing the aperture in the outer DC electrode minimizes its effect on the trapping RF field due to the greater distance from the ion. However, this placement requires a large angle ($\approx 70^{\circ}$) of the outcoupled beam. According to the Bragg condition for backward grating couplers, larger outcoupling angles correspond to higher angle-to-period sensitivity, making them more sensitive to fabrication tolerances. For example, a 200 nm thick Si$_3$N$_4$ backward grating coupler at a wavelength of 760 nm exhibits an angle-to-period sensitivity of 1.4°/nm at $\theta = -70$°, see figure \ref{fig:grating_coupler}(b). Given a typical fabrication tolerance of 5 nm \cite{bojko2011electron, Siew2021}, this results in an angular deviation of $7^{\circ}$. Combined with the increased distance from the target ion position, this deviation leads to significant beam misalignment of the laser beam to the ion.  
		Using forward grating couplers alleviates this issue by having smaller  angle-to-period sensitivity at the same angle, which are less sensitive to fabrication tolerances. However, forward grating couplers operated at large angles generate higher-order beams, as shown in figure \ref{fig:grating_coupler}. These unintended beams contribute to stray light and optical cross-talk,  introducing errors that degrade the performance of atomic clocks and quantum logic operations. Although the apertures in the center DC and RF electrodes displace the ion in opposing directions due to the sign change and the varying dominance of distorted RF field components, their effects on the trapping field are comparable. Therefore, in the subsequent sections, we focus on the aperture in the RF electrode.

		While excess micromotion arising from a displacement of the RF field minimum can be compensated using static electric fields, the associated changes in the secular frequencies cannot be eliminated without perturbing the trapping potential. As shown in figure \ref{fig:compare_minimum_Erf_single_hole}, the displacement of the $E_\mathrm{rf,r}$ minimum is nevertheless strongly correlated with the underlying distortion of the RF pseudopotential and with the corresponding changes in the secular frequencies. We therefore use the displacement of the $E_\mathrm{rf,r}$ minimum and the residual RF field along the trap axis as the primary indicators of aperture-induced distortion. An additional motivation for this choice stems from the integrated-optics approach, in which the optical beams are aligned to the geometric trap center. In such configurations, positioning the ion exactly at the RF-field minimum may not always be practical without degrading the optical overlap. 
		The dependence of the secular frequencies on the aperture width, where significant variations can occur, is discussed explicitly in the corresponding section.

		\subsection{Effect of aperture position and geometry}

		\subsubsection{Effect of aperture position}

		\noindent{In this section, we examine how the field distortion scales with the position of the aperture in the RF electrode. We first vary the position of an aperture in z-direction $p_\mathrm{z}$	
			
			\begin{figure}[!htbp]
				
				\centering
				\includegraphics[scale=0.36]{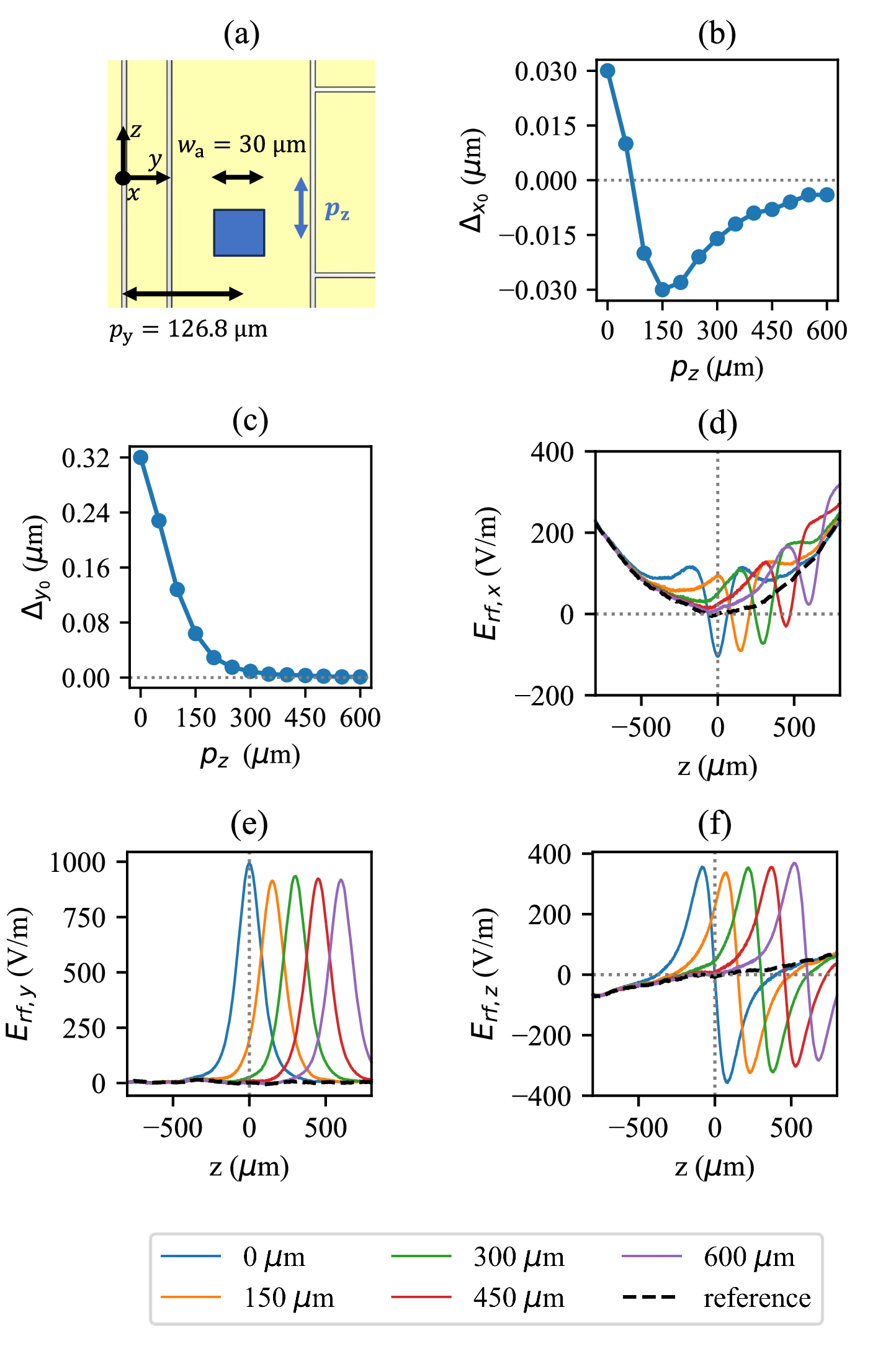}
				\caption{ (a) Varying $p_\mathrm{z}$ of an aperture in the RF electrode, with $w_\mathrm{a}$ = 30 $\mathrm{\mu}$m and $p_\mathrm{y}$ = 126.8 $\mathrm{\mu}$m (centered in the RF electrode). Subplots (b) and (c) illustrate  the  displacement of the $E_\mathrm{rf,r}$ minimum in x-y plane along  x- and y-directions as a function of $p_\mathrm{z}$. Subplots (d)-(f) show the x-, y- and z-components of the RF field along the trap axis for $p_\mathrm{z}$ varying from  0 $\mathrm{\mu}$m to 600 $\mathrm{\mu}$m and the reference case (black dashed line).}
				\label{fig:pz}
			\end{figure}

			\noindent{in the RF electrode while keeping $w_\mathrm{a}$ = 30 $\mathrm{\mu}$m and $p_\mathrm{y}$ = 126.8 $\mathrm{\mu}$m (centered in the RF} electrode), as shown in figure \ref{fig:pz}. Figures \ref{fig:pz}(b) and (c) present the impact on the displacement of the   $E_\mathrm{rf,r}$ minimum in the x-y plane by varying $p_\mathrm{z}$  from 0 $\mathrm{\mu}$m to 600 $\mathrm{\mu}$m.}
		
		The displacement of the $E_\mathrm{rf,r}$ minimum  in x-direction ($\Delta_\mathrm{x_0}$) 
		shifts from positive to negative as $p_\mathrm{z}$  increases from 0 to  150 $\mathrm{\mu}$m, with the displacement vanishing at $p_\mathrm{z}$ = 75 $\mathrm{\mu}$m. As the aperture is placed farther from the trap center, the displacement approaches zero. In contrast, the displacement in y-direction ($\Delta_\mathrm{y_0}$) follows a monotonic trend, decreasing steadily and approaching zero as  $p_\mathrm{z}$ increases. 
		The analysis along the trap axis (figures \ref{fig:pz}(d)-(f)) demonstrates that the RF field maintains its characteristic peak shapes similar to that shown by the red line in figure \ref{fig:compare_all_axial} while undergoing a spatial shift that corresponds to the aperture's position. 	When the aperture is moved  farther away from the trap center, the amplitudes and gradients of the RF field at the target ion position converge to the case without any aperture (black dashed line), since the influence of the aperture on the RF field at the target ion position decreases.

		\begin{figure}[!htbp]
			
			\centering
			\includegraphics[scale=0.36]{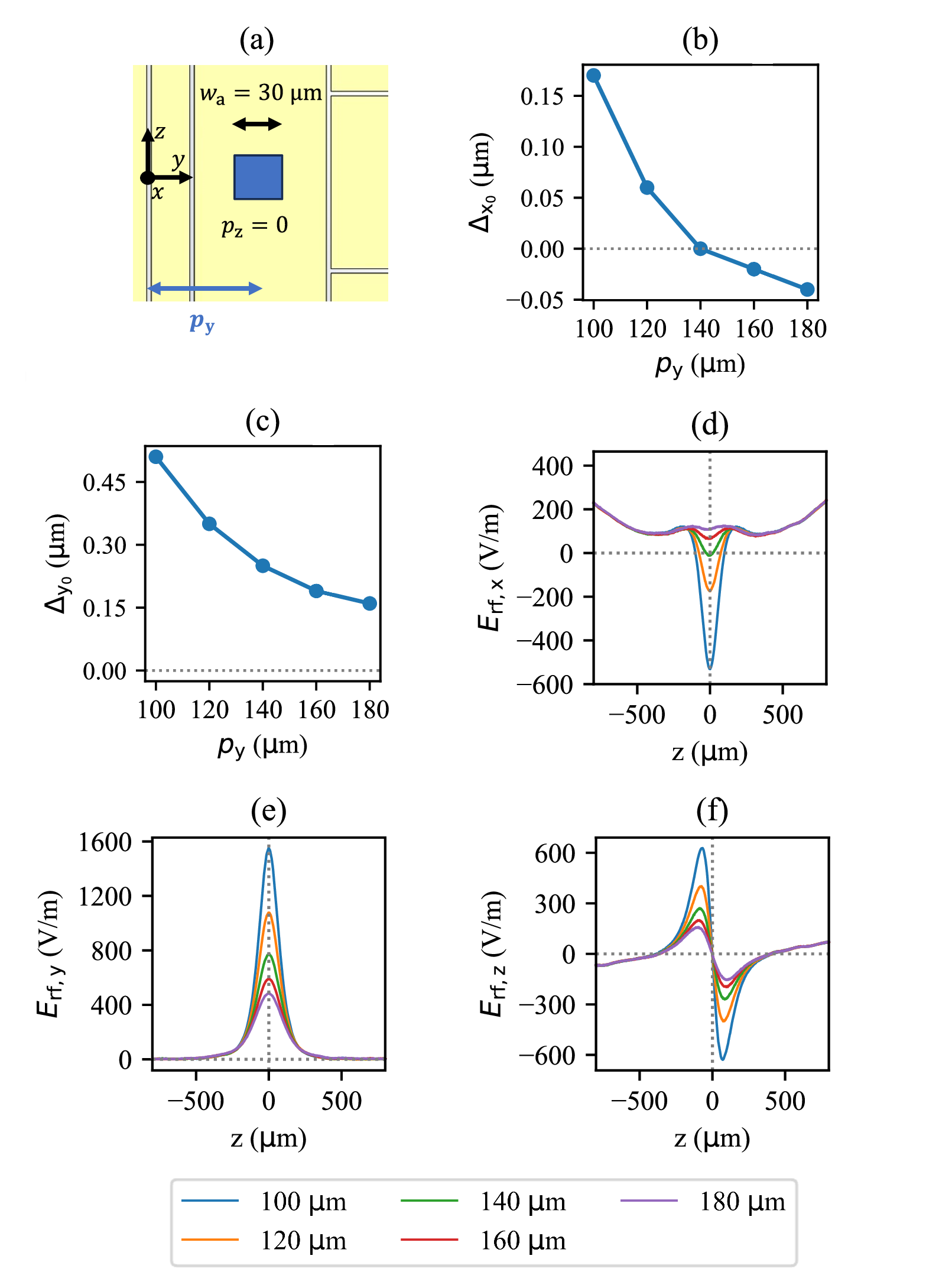}
			\caption{	 (a) Varying $p_\mathrm{y}$ of an aperture in the RF electrode in y-direction, with $w_\mathrm{a}$ = 30 $\mathrm{\mu}$m and $p_\mathrm{z}$ = 0 $\mathrm{\mu}$m. Subplots (b) and (c) illustrate the  displacement of the $E_\mathrm{rf,r}$ minimum in the x-y plane along  x- and y-directions as a function of $p_\mathrm{y}$. Subplots (d)-(f) show the x-, y- and z-components of the RF field along the trap axis  for
				$p_\mathrm{y}$ varying from  100 $\mathrm{\mu}$m to 180 $\mathrm{\mu}$m.}
			\label{fig:py}
		\end{figure}

		We then vary the position of an aperture in y-direction ($p_\mathrm{y}$) in the RF electrode while keeping   $w_\mathrm{a}$ = 30 $\mathrm{\mu}$m and  $p_\mathrm{z}$ = 0 $\mathrm{\mu}$m (on the y-axis), as shown in figure \ref{fig:py}. Figures \ref{fig:py}(b) and (c) illustrate the effect on the displacement of the $E_\mathrm{rf,r}$ minimum in  x-y plane by varying $p_\mathrm{y}$ from 100 $\mathrm{\mu}$m to 180 $\mathrm{\mu}$m. The displacement of the $E_\mathrm{rf,r}$ minimum  in x-direction ($\Delta_\mathrm{x_0}$) crosses from positive to negative as $p_\mathrm{z}$  increases, with the displacement vanishing at $p_\mathrm{z}$ = 140 $\mathrm{\mu}$m, while the  displacement in y-direction ($\Delta_\mathrm{y_0}$) follows a monotonic downward trend, approaching zero as $p_\mathrm{y}$ increases. Along the trap axis, as shown in figures \ref{fig:py}(d)-(f), the peaks in the x- and y-components of the residual field remain centered  at z = 0 and increase in magnitude as $p_\mathrm{y}$ decreases.  The z-component maintains its dispersive lineshape with a gradient at $z = 0$ increasing significantly as $p_\mathrm{y}$ decreases.
		
		To summarize the effect of the aperture position, simulations indicate that placing the aperture farther from the trap center reduces RF field distortion at the target ion position, as expected. However, this geometric relationship creates a trade-off: greater distances require larger outcoupling angles, which in turn demand wider apertures.

		
		\subsubsection{Effect of aperture size}

				\begin{figure}[t]
			
			\centering
			\includegraphics[scale=0.36]{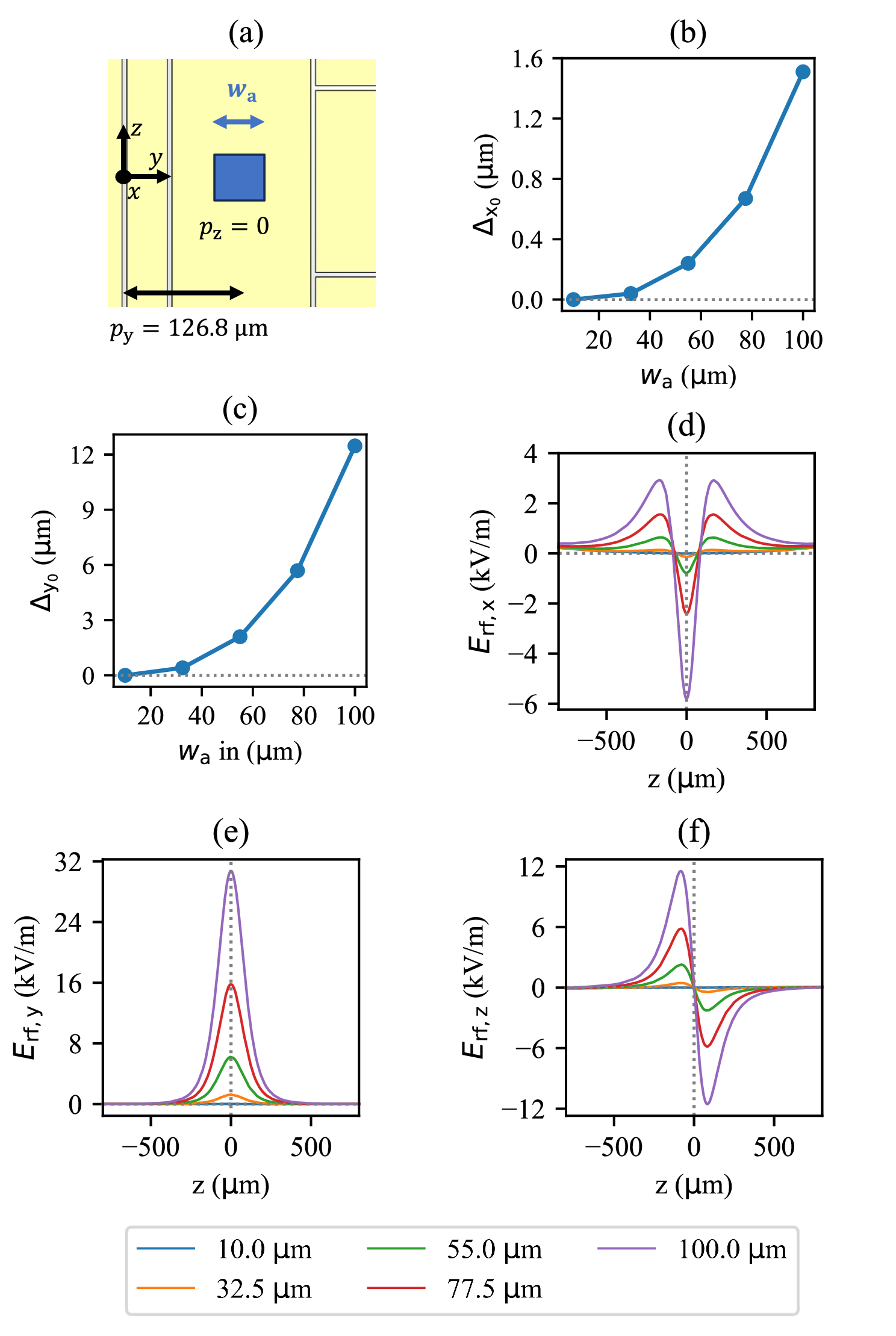}
			\caption{(a) Varying  $w_\mathrm{a}$ of an aperture in the RF electrode, with a fixed aperture position  $p_\mathrm{y}$ = 126.8 $\mathrm{\mu}$m and  $p_\mathrm{z}$ = 0 $\mathrm{\mu}$m. Subplots (b) and (c) illustrate  the  displacement of the $E_\mathrm{rf,r}$ minimum in x-y plane along  x- and y-directions as a function of $w_\mathrm{a}$. Subplots (d)-(f) show the x-, y- and z-components of the RF field along the trap axis  for $w_\mathrm{a}$ varying from 10 $\mathrm{\mu}$m to 100 $\mathrm{\mu}$m.}
			\label{fig:wa}
		\end{figure}
		
								\begin{figure}[h]
	
	\centering
	\includegraphics[scale=0.5]{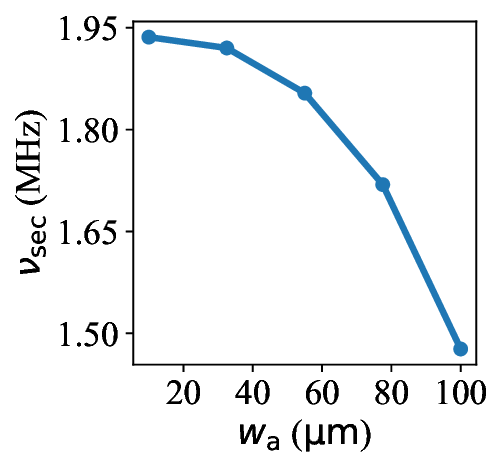}
	\caption{Radial secular frequency $\nu_\mathrm{sec}$ for a $^{172}\mathrm{Yb}^+$ ion as a function of the aperture width $w_\mathrm{a}$ for an aperture in the RF electrode. 
		The aperture position is fixed at $p_\mathrm{y}=126.8~\mu\mathrm{m}$ and $p_\mathrm{z}=0~\mu\mathrm{m}$.
	}
	\label{fig:secular_frequency_width}
\end{figure}
		
		\noindent{The required aperture size for the outcoupled beam depends on the ion height, the angle of the outcoupled beam, the required beam waist at the target ion position and the fabrication tolerances. The ion traps with integrated waveguides and grating couplers shown so far in reference \cite{Mehta2016,Niff2020, Ivory2021, Kwon2024, Mehta2020, Mordi2025} have   aperture sizes ranging  from 10 $\mathrm{\mu}$m to 35 $\mathrm{\mu}$m, with ion heights varying from 20 $\mathrm{\mu}$m to 55 $\mathrm{\mu}$m.  In our study, due to our greater ion height of 100 $\mathrm{\mu}$m,  we examine RF field distortion by varying aperture width across a broader range from 10 $\mathrm{\mu}$m to 100 $\mathrm{\mu}$m while maintaining a fixed position in the RF electrode, as shown in figure \ref{fig:wa}.

			 Increasing the aperture size from 10 $\mathrm{\mu}$m to 100 $\mathrm{\mu}$m causes a displacement of the $E_\mathrm{rf,r}$ minimum  in x-y plane by up to 1.6 $\mathrm{\mu}$m  in  x-direction and  up to 12 $\mathrm{\mu}$m in y-direction. Along the trap axis, the peaks of the x- and y-components of the residual field remain centered  at z = 0 and become significantly larger  as $w_\mathrm{a}$ increases. The x-component reaches an amplitude of approximately 6 kV/m at z = 0 with a strong gradient of 64.2 $\mathrm{V/mm^2}$ at the FWHM, while the y-component reaches  about 30 kV/m at z = 0 with a gradient of 163.3 $\mathrm{V/mm^2}$ at the FWHM. The z-component  maintains its characteristic dispersive lineshape with the gradient at z = 0 increasing significantly as $w_\mathrm{a}$ increases,  reaching 139 $\mathrm{V/mm^2}$ for $w_\mathrm{a}$  = 100 $\mathrm{\mu}$m.

			\begin{figure}[t]
				
				\centering
				\includegraphics[scale=0.36]{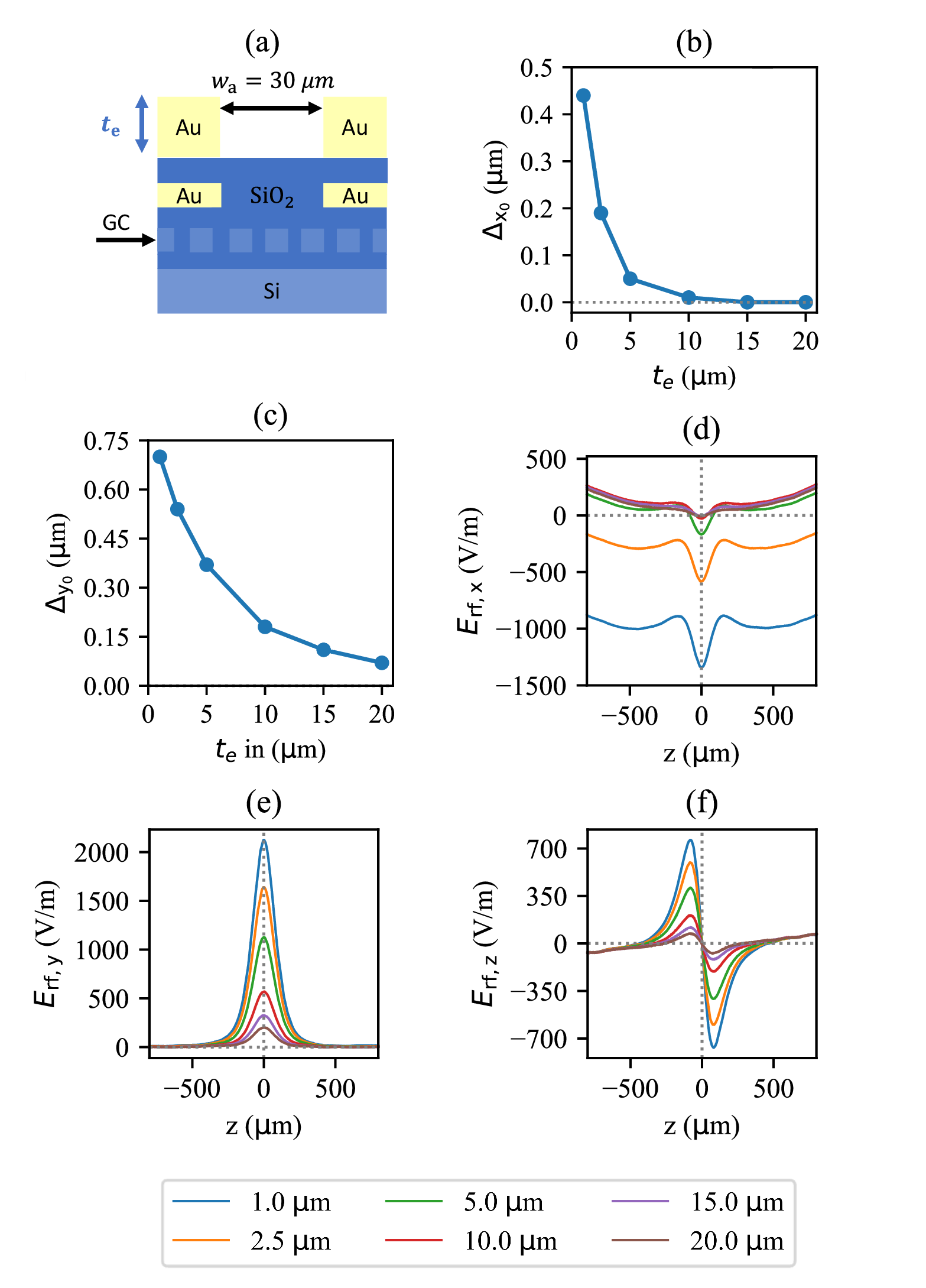}
				\caption{(a) Varying $t_\mathrm{e}$ of the electrode around an aperture in the RF electrode, with $w_\mathrm{a}$ = 30 $\mathrm{\mu}$m and $p_\mathrm{y}$ = 126.8 $\mathrm{\mu}$m,   $p_\mathrm{z}$ = 0 $\mathrm{\mu}$m. The layer-stack is schematic and not drawn to scale. Subplots (b) and (c) illustrate  the  displacement of the $E_\mathrm{rf,r}$ minimum in x-y plane along  x- and y-directions as a function of $t_\mathrm{e}$. Subplots (d)-(f) show the x-, y- and z-components of the RF field along the trap axis for $t_\mathrm{e}$ varying from 1 $\mathrm{\mu}$m to 20 $\mathrm{\mu}$m.}
				\label{fig:te}
			\end{figure}
			
			We further investigate the change in the secular frequency as a function of the aperture width.
			As shown in figure \ref{fig:secular_frequency_width}, increasing $w_\mathrm{a}$ leads to a monotonic reduction of the radial secular frequency. 
			For small apertures, this effect remains weak: for the reference aperture width $w_\mathrm{a}=30~\mu\mathrm{m}$ used throughout this work, the secular-frequency change is limited to about $0.7\%$. 
			With increasing aperture width, however, the curvature distortion becomes substantial. 
			In particular, increasing the aperture width from $w_\mathrm{a}=30~\mu\mathrm{m}$ to $100~\mu\mathrm{m}$ results in a reduction of the radial secular frequency of up to approximately $20\%$.

			In comparison to variations in location, altering the size of the aperture has a more pronounced effect on the electric field distortion. A larger distance of the aperture from the trap center reduces distortion but requires a larger aperture, which in turn introduces additional distortion. Therefore, a careful balance must be struck between the distance and the size of the aperture.

			\subsubsection{Effect of electrode thickness}

			\noindent{Next, we study the  effect of the electrode thickness on the field distortion as presented in figure \ref{fig:te}.} Our results demonstrate that increasing electrode thickness reduces both the displacement of the $E_\mathrm{rf,r}$ minimum in x-y plane and the residual RF field along the trap axis. This improvement is due to the reduction of the edge effects. Edge effects arise when the geometry of the electrode, such as sharp edges or boundaries, introduces non-uniformities in the electric field, especially near the edges. The aperture in a thinner electrode has a more pronounced field distortion at its edges, because the field lines are less confined and can spread outward more easily, leading to larger variations in the trapping field. This behavior is similar to that seen in larger capacitors, where the influence of edge effects is diminished as the length of the capacitor (or in our case, the thickness of the electrode) increases. However, increasing the electrode thickness also imposes geometric constraints, as excessively thick electrodes may obstruct the outcoupled laser beam, leading to a trade-off between reduced field distortions and optical access.

			\subsection{Effect of symmetry}
			
			When a single aperture is positioned off the symmetry axes (either y or z), as illustrated in figure \ref{fig:symmetry}(a), the $E_\mathrm{rf,r}$ minimum in x-y plane is displaced from its target position. The asymmetry also leads to residual RF field amplitude and gradients in all three components at the target ion position.

			\begin{figure}[b]
				
				\centering
				\includegraphics[scale=0.34]{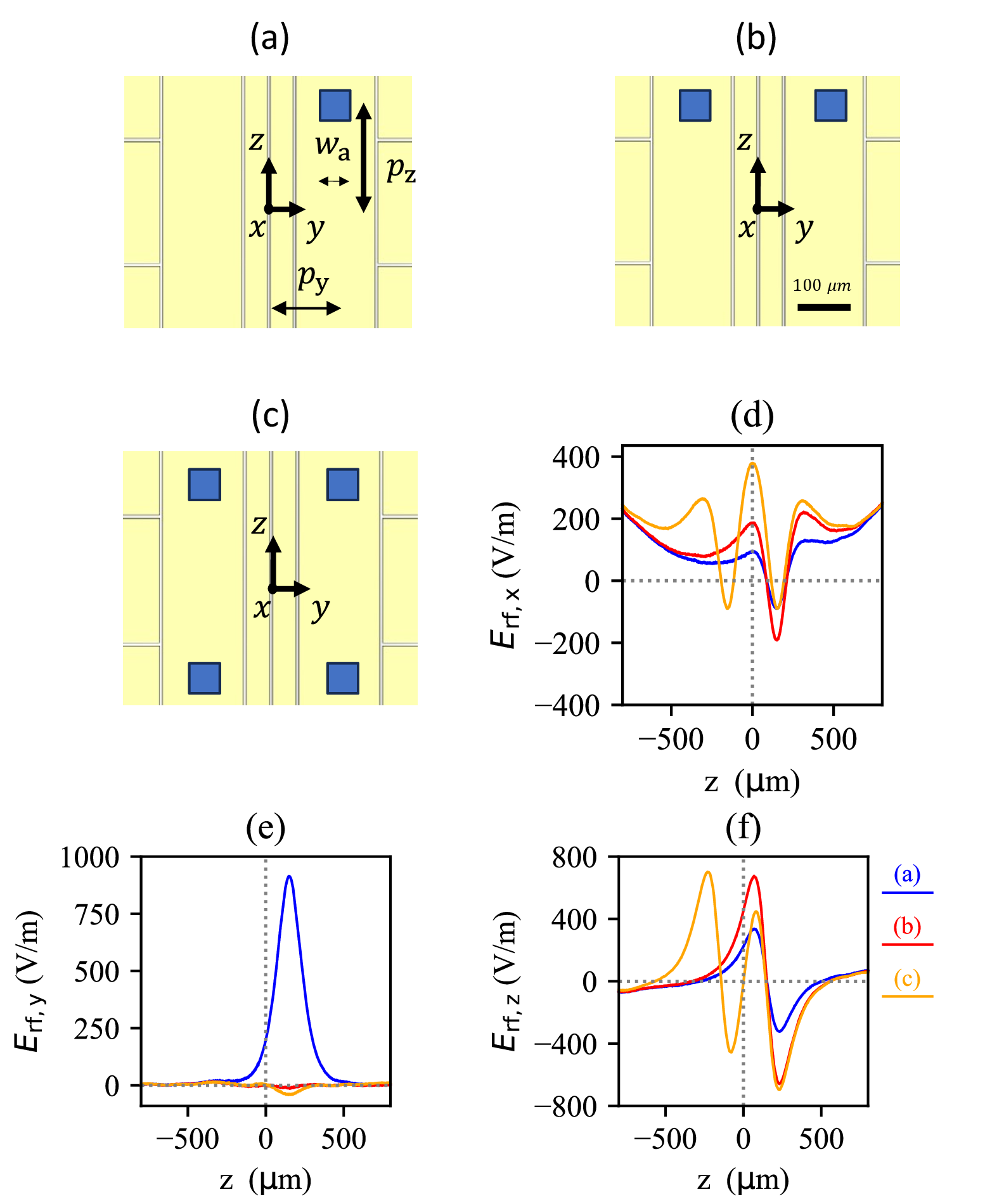}
				\caption{(a) A single square-shaped aperture positioned asymmetrically, with a width  of $w_\mathrm{a}$ = 30 $\mathrm{\mu}$m, located at  $p_\mathrm{y}$ = 126.8 $\mathrm{\mu}$m,   $p_\mathrm{z}$ = 100 $\mathrm{\mu}$m. (b) Configuration obtained by mirroring the aperture in subplot (a) with respect to the z-axis.	(c) Configuration obtained by mirroring both apertures in subplot (b) with respect to the y-axis. Subplots (d)-(f) illustrate  the x-, y- and z-components of the RF field along the trap axis, respectively. Blue, red and orange lines indicate the configurations in subplots (a)-(c), respectively.}
				
				\label{fig:symmetry}
				
			\end{figure}

			To address the displacement in the radial direction, we first introduce an additional aperture placed symmetrically with respect to the z-axis, as shown in figure \ref{fig:symmetry}(b). The analysis along the trap axis (figures \ref{fig:symmetry}(d)-(f), red line) demonstrates that this z-symmetry cancels the y-component of the RF field along the entire trap axis, while simultaneously enhancing the peaks of the x- and z-components.
			
			Next, to compensate the residual RF amplitude in the z-component at the target ion position (z = 0), we introduce two additional apertures to create y-symmetry, as depicted in figure \ref{fig:symmetry}(c). The analysis along the trap axis (figures \ref{fig:symmetry}(d)-(f), orange line) shows that while $E_\mathrm{rf,z}$ vanishes at z = 0 as expected, the introduction of the extra apertures creates additional peaks in the field. Furthermore, the x-component exhibits a more complex lineshape with additional peaks and a non-vanishing field amplitude at z = 0. Since surface ion traps are inherently two-dimensional, exploiting x-symmetry for further compensation is not feasible, meaning that the x-component cannot be fully mitigated in this configuration.

			\subsection{Effect of using transparent conductive oxide coating}
			
			The use of transparent conductive oxides (TCOs) \cite{Stad2012} in ion-trap experiments has been demonstrated in several recent works \cite{Niff2020, xing2025rapidmultimodetrappedionlaser, wang2025tcostransformcavityqed}. TCOs  can be applied as thin film coatings to cover the aperture, as shown in figure \ref{fig:cut_view_ITO}. These thin films are electrically connected to the electrodes, thus sharing their potential while allowing outcoupled light to pass through. In addition, the TCO covers dielectrics that could potentially charge up due to laser-induced charging \cite{wang2011,Ong2020} and  detrimentally affect the electric field at the ion location. A widely used TCO is indium tin oxide (ITO). 
			In ITO films, the thickness is limited by a trade-off between electrical conductivity and optical transmittance. Increasing the thickness improves the conductivity but reduces the transmission due to increased absorption. For a 51\,nm thick ITO layer, Jansson \textit{et al.} \cite{Jansson2024} report an DC electrical conductivity at room temperature of approximately $1.7 \times 10^{5}\,\mathrm{S/m}$, while maintaining high optical transmission of about 66\,\% for a single ITO layer and about 80\,\% when combined with an anti-reflection coating at 370\,nm. 
			At longer visible and near-infrared wavelengths, such as 760\,nm and 935\,nm, the optical transmittance exceeds 95\,\%.
			Accordingly, we assume a TCO layer thickness of $t_{\mathrm{TCO}} = 50\,\mathrm{nm}$ in all subsequent simulations.

			\begin{figure}[b]
				
				\centering
				\includegraphics[scale=0.06]{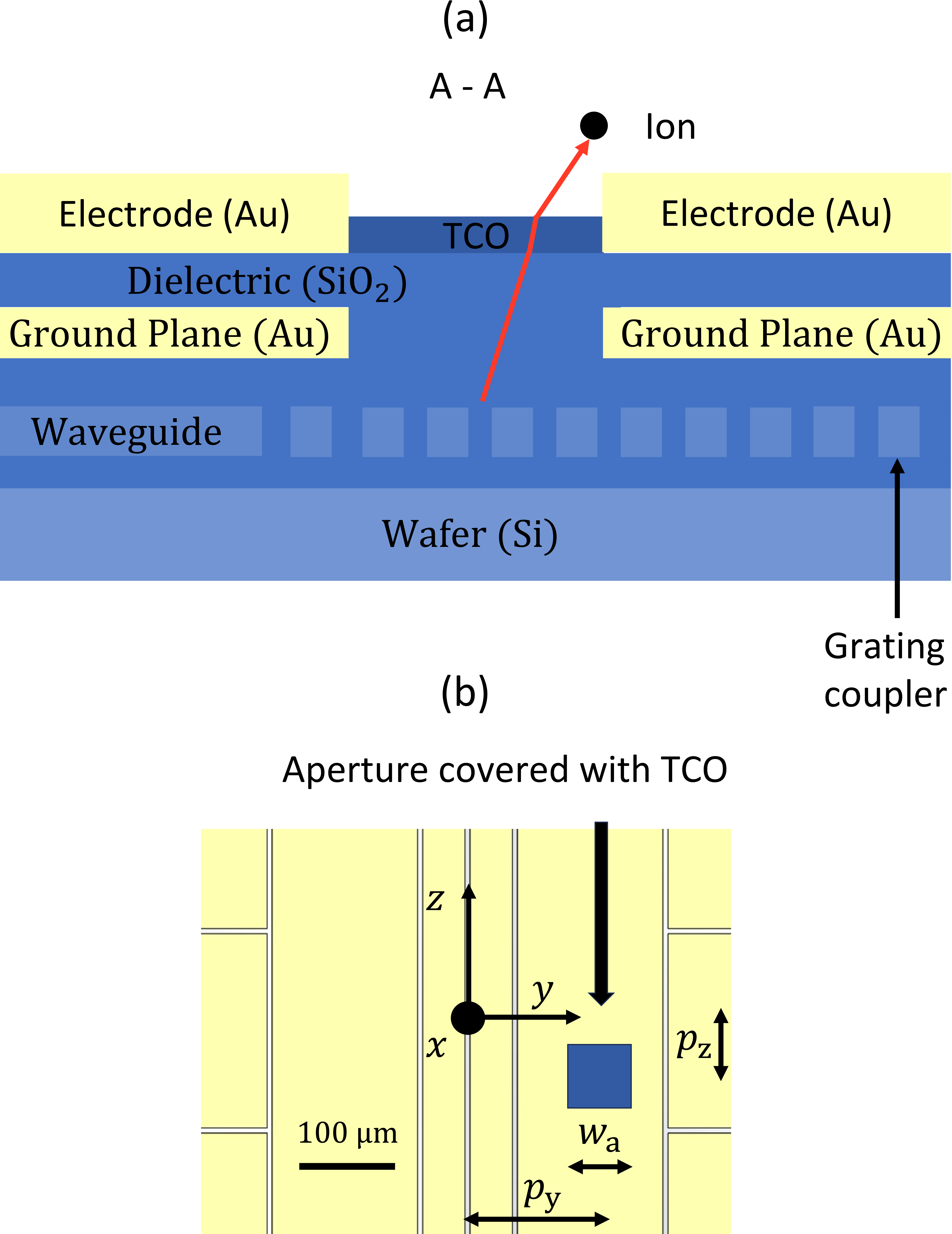}
				\caption{(a) A cross-sectional view (A-A) illustrates that a transparent and conductive coating is applied to cover an aperture in the electrode. The cross-sectional view is schematic and not drawn to scale. (b)  Top view of an aperture coated with TCO with a thickness of 50 nm in the gold electrode with  $t_\mathrm{e}$ = 6 $\mathrm{\mu}$m.}
				\label{fig:cut_view_ITO}
			\end{figure}
			
			\subsubsection{Effect of modified topography of the electrode}\label{Effect of modified topography of the electrode}\hfill
			
			\noindent{To account for the thickness difference in the trap electrode at the outcoupling aperture, we initially assume an idealized TCO layer with a conductivity equivalent to gold (4.5 × $10^7$ S/m), thereby ensuring that the layer shares the same electrical potential as the gold electrode. }

			\begin{figure}[t]

				\centering
				\includegraphics[scale=0.45]{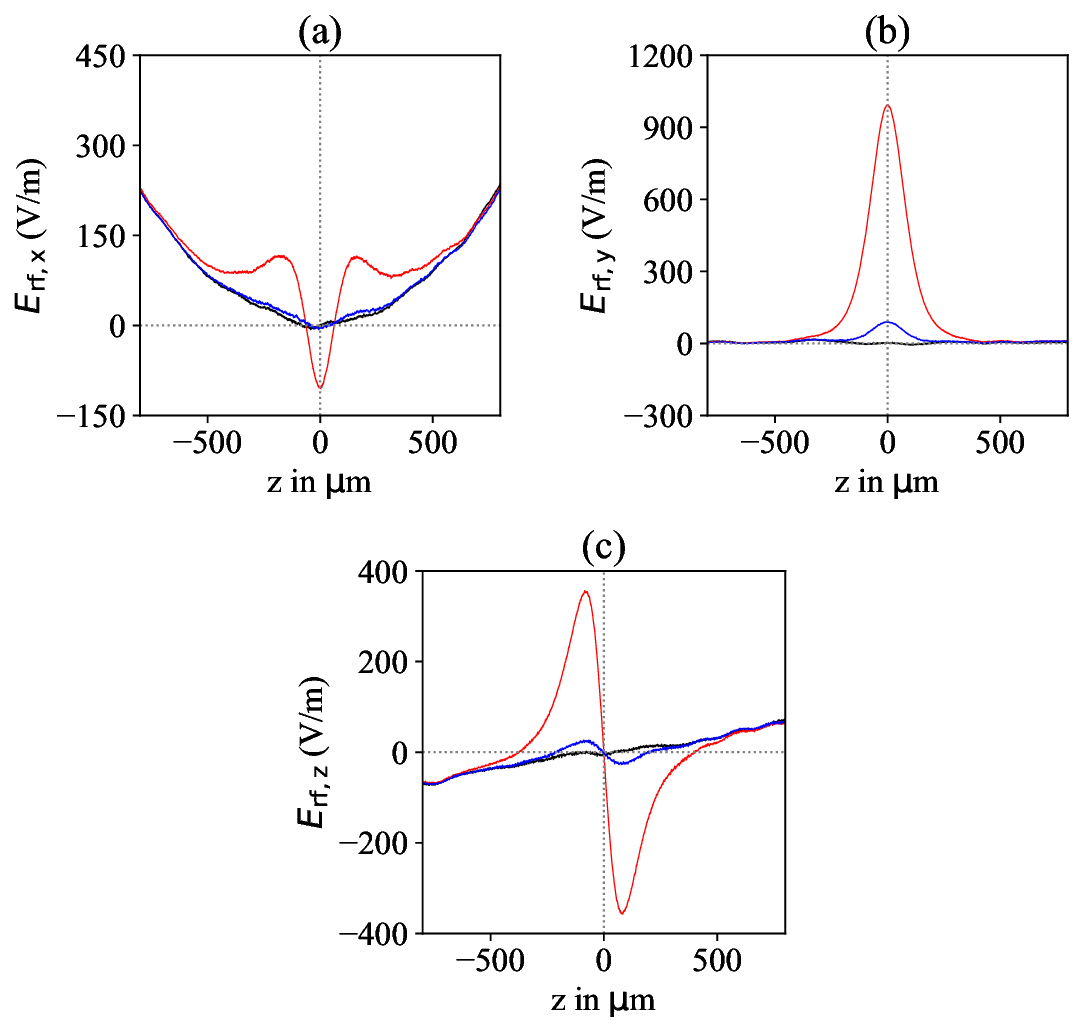}
				\caption{ Subplots (a)-(c) illustrate  the x-, y- and z-components of the RF field along the trap axis: reference configuration (black line), a single square-shaped aperture with $w_\mathrm{a}$ = 30 $\mathrm{\mu}$m, $p_\mathrm{y}$ = 126.8 $\mathrm{\mu}$m, and $p_\mathrm{z}$ = 0 $\mathrm{\mu}$m in the RF electrode without TCO coating (red line) and the same aperture with idealized TCO coating of $t_{\mathrm{TCO}}$ = 50 nm and conductivity equal to gold (blue line).}
				\label{fig:compare_all_axial_for_TCO}
			\end{figure}
			
			The simulation results (see figure \ref{fig:compare_all_axial_for_TCO}) indicate that, although the RF field along the trap axis are substantially reduced when TCO is applied (blue line) in comparison to the case without TCO (red line), a small residual RF field remains in comparison to the reference configuration (black). In the radial direction,  the $E_\mathrm{rf,r}$ minimum in x-y plane exhibits a residual displacement of approximately 30 nm in y-direction and no displacement in x-direction when the aperture is covered with TCO. Although the displacements are significantly smaller than in the case without the TCO coating (30 nm in x-direction and 320 nm in y-direction as shown in figure \ref{fig:compare_minimum_Erf_single_hole}(c)), they do not vanish entirely.}

		\subsubsection{Effect of varying the conductivity of TCO}\label{Effect of varying the conductivity of TCO}\hfill

		\begin{figure}[!htbp]
			
			\centering
			\includegraphics[scale=0.36]{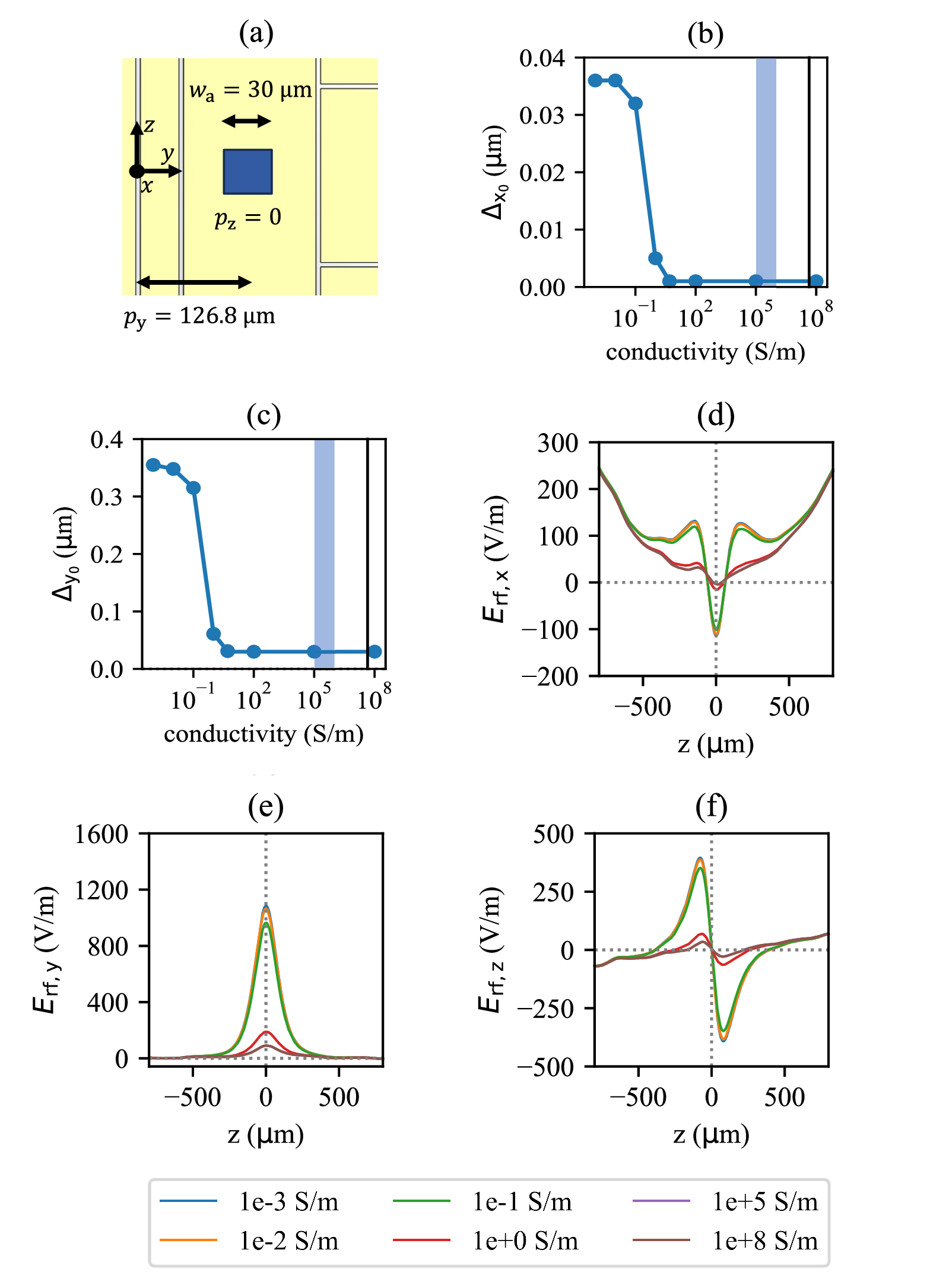}
			\caption{(a) Variation of the conductivity of the TCO covering an aperture with a fixed width of $w_\mathrm{a}$ = 30 $\mathrm{\mu}$m and a constant  position of $p_\mathrm{y}$ = 126.8 $\mathrm{\mu}$m and   $p_\mathrm{z}$ = 0 $\mathrm{\mu}$m in the RF electrode. Subplots (b) and (c) illustrate the displacement of the $E_\mathrm{rf,r}$ minimum in x-y plane along  x- and y-directions as a function of the TCO conductivity, the black line indicates the conductivity of gold, the blue band spans the region of typical conductivities of ITO.  Subplots (d)-(f) show the x-, y- and z-components of the RF field along the trap axis for a conductivity from $10^{-3}$ S/m to $10^8$ S/m. Notably, the purple and brown lines overlap.}
			\label{fig:ITO_phase_assembly}
			
		\end{figure}

		\noindent{The conductivity of TCO is strongly influenced by its material composition and manufacturing process. To assess the effect of the TCO conductivity on RF field distortion, we vary the conductivity of the TCO layer and model the RF field by taking into account the current flow through both the gold electrode and the TCO. This simulation cannot be treated statically, therefore we use the electric current module in COMSOL. An voltage source with a frequency of 16 MHz and an amplitude of 100 V is applied to the RF electrodes. All other DC electrodes are grounded. This method captures both the phase and potential distortions in the electrode, from which the RF field at the target ion position is calculated.}

		In x-y plane, the displacement of the $E_\mathrm{rf,r}$ minimum  decreases in both x- and y-directions as the conductivity of TCO increases, as shown in figure \ref{fig:ITO_phase_assembly}. The blue band indicates the region of typical conductivities of ITO and the black line indicates the conductivity of gold. When the conductivity of TCO matches that of gold ($4.5 \times 10^7$ S/m), the displacement in x-direction vanishes, while a residual displacement of 30 nm remains in y-direction, which is consistent with results from the electrostatic simulation in Section \ref{Effect of modified topography of the electrode}. A similar trend is observed in the axial direction as well. As the conductivity of the TCO increases to match that of gold, the RF field along the trap axis converges to the results shown by the blue line in figure \ref{fig:compare_all_axial_for_TCO}.

		\subsubsection{Effect of the RF phase shift}\label{Effect of the RF phase shift}\hfill

		\noindent{The aperture in the RF electrode not only distorts the potential of the electrode, but also introduces a phase delay due to the differences in conductivity and thickness between gold and TCO. Such phase difference between the two RF electrodes can cause a non-zero  $E_\mathrm{rf,r}$ minimum, resulting in  excess micromotion that can not be compensated \cite{Berkeland1998}.}

		\begin{figure}[b]
			
			\centering
			\includegraphics[scale=0.8]{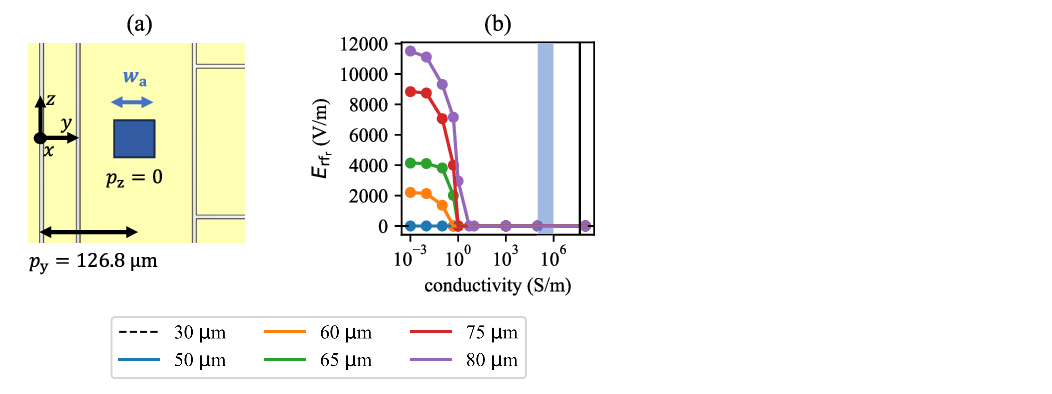}
			\caption{(a) Variation of the conductivity of the TCO covering an aperture with varying widths ($w_\mathrm{a}$).  The position of the aperture is kept constant at $p_\mathrm{y}$ = 126.8 $\mathrm{\mu}$m and  $p_\mathrm{z}$ = 0 $\mathrm{\mu}$m in the RF electrode. (b) Amplitude of $E_\mathrm{rf,r}$ minimum in x-y plane as a function of the conductivity of TCO for aperture with widths of $w_\mathrm{a}$ varying from  30 $\mathrm{\mu}$m to 80 $\mathrm{\mu}$m. All lines converge to zero and overlap when the TCO conductivity exceeds 10 S/m. The black line indicates the conductivity of gold, the blue band  spans the region of typical conductivities of ITO.}
			\label{fig:radial_ITO_phase_Erf_minimum_norm}
			
		\end{figure}
		
		Figure \ref{fig:radial_ITO_phase_Erf_minimum_norm} shows the residual amplitude of  $E_\mathrm{rf,r}$ at its minimum as a function of the TCO conductivity for varying aperture widths $w_\mathrm{a}$. The residual amplitude exhibits distinct behavior across different conductivity regimes. At low conductivities (below 1 S/m) with $w_\mathrm{a}>$ 50 $\mathrm{\mu}$m,  $E_\mathrm{rf,r}$   exhibits a significant residual amplitude at its minimum. This arises from the mismatch in conductivity and thickness between the gold and TCO coatings, which induces a phase delay in the RF electrode and results in a non-vanishing field amplitude at the $E_\mathrm{rf,r}$ minimum. As the conductivity decreases further to $10^{-3}$ S/m, where the aperture effectively behaves as if it had no TCO coating, the residual amplitude converges to a maximum. This indicates that the cut-through in the RF electrode created by the aperture  introduces strong phase distortions (for $w_\mathrm{a}>$ 50 $\mathrm{\mu}$m), resulting in a residual  $E_\mathrm{rf,r}$ minimum that can not be fully compensated. Conversely, increasing TCO conductivity reduces the residual field amplitude.  When the conductivity reaches the value of commonly used ITO or matches that of gold, the residual amplitude vanishes for all aperture widths. This indicates that the phase delay due to thickness differences between the TCO and gold is negligible.

		To summarize the effect of TCO,	our findings indicate that the effect of the conductivity of the TCO on the RF field can be neglected over a large range from 10 S/m to $10^{8}$ S/m.  The observed conductivity threshold on the order of $10~\mathrm{S/m}$ may be qualitatively interpreted by modeling the ITO patch as a resistance element embedded in a  capacitive network formed by the surrounding gold electrodes and dielectric layers. 
		Whether the ITO patch follows the RF potential of the surrounding gold electrode is governed by the effective charging dynamics of the ITO patch. 
		For lower conductivity, this charging time becomes longer than the RF period, such that the patch cannot fully follow the RF drive, whereas for sufficiently high conductivity the patch potential approximately follows the RF drive, thereby suppressing field distortions. 
		However, a detailed quantitative analysis of the underlying impedance network is beyond the scope of the present work.
		The conductivity and the thickness of the commonly used ITO layers are adequate to mitigate significant RF field distortion without introducing a phase shift in the RF electrode. However, the topography of the surface electrodes with apertures prevents a complete elimination of the RF field distortion.


		\section{Conclusion and Outlook}

		In this paper, we present a comprehensive analysis of the effects on the trapping field of a photonic-integrated surface ion trap caused by apertures for vertically outcoupled laser beams from waveguides. We find that placing apertures in the outer DC electrode is the optimal strategy to avoid significant trapping field distortion. However, this requires larger outcoupling angles. For backward grating couplers, larger angles necessitate smaller grating periods and have higher angle-to-period sensitivity, making them highly sensitive to fabrication tolerances. In the case of forward grating couplers, for larger angles higher-order diffraction are coupled out. These effects become more pronounced at shorter wavelengths. For instance, given a typical fabrication tolerance of 5 nm, a 120 nm thick Al$_2$O$_3$ backward grating coupler at a wavelength of 370 nm exhibits an angle-to-period sensitivity of 2.7°/nm at $\theta = -70$°, leading to an angular deviation of approximately 13.5°. Such a grating coupler requires a feature size of approximately 70 nm, which is critical for foundries \cite{Siew2021}.

		When the aperture is located in the RF electrode, a trade-off emerges between the distance from the trap center and the size of the aperture. As the distance $p_y$ increases, the distortion of the trapping field leads to a smaller displacement of the $E_\mathrm{rf,r}$ minimum. For distances from $p_y=65$ $\mu$m to $p_y=180$ $\mu$m in the RF electrode, the displacement reduces from $\Delta_{\mathrm{y0}}=1.5$ $\mu$m to 150 nm in the y-direction and $\Delta_{\mathrm{x0}}=1.5$ $\mu$m  to 50 nm in the x-direction. For a shift of the aperture position in the RF electrode in the z-direction, a similar behavior is observed with a displacement on the same scale. However, the distortion and the resulting displacement also depend on the aperture size. While the displacement is $<1$ $\mu$m for a square-shaped aperture of width $w_a=30$ $\mu$m at $p_\mathrm{y}$ = 126.8 $\mathrm{\mu}$m and  $p_\mathrm{z}$ = 0 $\mathrm{\mu}$m, for a larger aperture with $w_a=100$ $\mu$m at the same position the ion is displaced by $\Delta_{\mathrm{x0}}=1.6$ $\mu$m and $\Delta_{\mathrm{y0}}=12$ $\mu$m. This large displacement is due to the highly distorted field,   with the y-component exhibiting a peak amplitude  of 30 kV/m and a gradient of 164 V/mm$^2$ at the target ion position. Importantly, the aperture should not be placed at the edge of the gap between the center DC and RF electrodes, as this results in a larger ion displacement of 1.5 $\mu$m in both directions compared to when the aperture is fully surrounded by the metal electrode. When the aperture is positioned in the center DC electrode, the ion displacement $\Delta_{\mathrm{x0}}$ can be minimized to 150 nm and $\Delta_{\mathrm{y0}}$ can be zeroed by placing the aperture directly at the trap center.

		\begin{table*}[t]
			\centering
			
			\setlength{\tabcolsep}{3pt}
			
			\caption{Electric field components, RF-minimum displacement at $z=0$, 
				micromotion-induced fractional time-dilation shift 
				$\left| \Delta\nu/\nu_0 \right|_{\mathrm{TD}}$, and RF-noise heating rates along the three motional
				directions, evaluated at the target ion position 
				$(x=100\,\mu\mathrm{m},\, y=0,\, z=0)$. 
				The position of the single aperture in the RF electrode 
				is fixed at $p_y = 126.8~\mu\mathrm{m}$ and $p_z = 0~\mu\mathrm{m}$. 	Four configurations are considered: 
				(i) a single-aperture geometry; 
				(ii) a single aperture with an ITO coating; 
				(iii) a symmetric two-aperture geometry obtained by introducing a second aperture 
				mirror-symmetric with respect to the $z$ axis; and 
				(iv) the symmetric configuration with ITO coating. All calculations assume a single $^{172}\mathrm{Yb}^+$ ion with
				$\nu_{\mathrm{sec},r} = 1.9\,\mathrm{MHz}$ and
				$\nu_{\mathrm{sec},z} = 300\,\mathrm{kHz}$.}
			
			\label{tab:dup_blocks_z=0}
			
			\begin{tabular}{|p{2.0cm}|c|p{1.5cm}|p{1.5cm}|c|c|c|c|c|c|c|}
				\hline
				\multicolumn{2}{|c|}{} 
				& \multicolumn{2}{c|}{RF min.\ disp.\ ($z=0$)} 
				& \multicolumn{7}{c|}{Evaluated at the target ion position} \\
				\hline
				\centering Configuration
				& \shortstack{$w_a$ \\ ($\mu$m)}
				& \centering \shortstack{$\Delta x_0$ \\ ($\mu$m)}
				& \centering \shortstack{$\Delta y_0$ \\ ($\mu$m)}
				& \shortstack{$E_x$ \\ (V/m)}
				& \shortstack{$E_y$ \\ (V/m)}
				& \shortstack{$E_z$ \\ (V/m)}
				& $\left| \Delta\nu/\nu_0 \right|_{\mathrm{TD}}$
				& \shortstack{$\dot{\bar n}_x$ \\ (s$^{-1}$)}
				& \shortstack{$\dot{\bar n}_y$ \\ (s$^{-1}$)}
				& \shortstack{$\dot{\bar n}_z$ \\ (s$^{-1}$)} \tabularnewline
				\hline
				
				\multirow{3}{*}{\centering single}
				& 30   & \centering 0.03 & \centering 0.32  & -103.3  & 993.3   & 0  & $8.6\times10^{-17}$ & $7.0\times10^{-2}$ & $6.3\times10^{0}$ & 0 \tabularnewline
				& 77.5 & \centering 0.68 & \centering 5.69  & -2402.5 & 15796.4 & 0  & $2.2\times10^{-14}$ & $6.3\times10^{1}$ & $1.3\times10^{3}$ & 0 \tabularnewline
				& 100  & \centering 1.50 & \centering 12.47 & -5794.8 & 30722.4 & 0  & $8.5\times10^{-14}$ & $5.0\times10^{2}$ & $4.2\times10^{3}$ & 0 \tabularnewline
				\hline
				
				\multirow{3}{*}{\centering \shortstack{single \\ + ITO}}
				& 30   & \centering 0.00 & \centering 0.03  & -4.2   & 89.3   & 0  & $6.9\times10^{-19}$ & $8.2\times10^{-5}$ & $5.1\times10^{-2}$ & 0 \tabularnewline
				& 77.5 & \centering 0.05 & \centering 0.32  & -154.1 & 978.0  & 0  & $8.5\times10^{-17}$ & $1.6\times10^{-1}$ & $6.1\times10^{0}$ & 0 \tabularnewline
				& 100  & \centering 0.13 & \centering 0.64  & -397.3 & 1960.7 & 0  & $3.5\times10^{-16}$ & $1.1\times10^{0}$ & $2.4\times10^{1}$ & 0 \tabularnewline
				\hline
				
				\multirow{3}{*}{\centering symmetry}
				& 30   & \centering 0.07 & \centering 0.01  & -212.7  & 0   & 0  & $4.1\times10^{-18}$ & $2.7\times10^{-1}$ & $0$ & 0 \tabularnewline
				& 77.5 & \centering 1.74 & \centering -0.01 & -4790.1 & 0   & 0  & $2.0\times10^{-15}$ & $1.2\times10^{2}$ & $0$ & 0 \tabularnewline
				& 100  & \centering 4.66 & \centering 0.08  & -11564.9 & 0   & 0  & $1.2\times10^{-14}$ & $6.3\times10^{2}$ & $0$ & 0 \tabularnewline
				\hline
				
				\multirow{3}{*}{\centering \shortstack{symmetry \\ + ITO}}
				& 30   & \centering 0.01 & \centering 0.01  & -16.5  & 0   & 0  & $7.2\times10^{-20}$ & $7.3\times10^{-4}$ & $0$ & 0 \tabularnewline
				& 77.5 & \centering 0.01 & \centering 0.01  & -304.3 & 0   & 0  & $8.2\times10^{-18}$ & $5.5\times10^{-1}$ & $0$ & 0 \tabularnewline
				& 100  & \centering 0.01 & \centering 0.01  & -804.7 & 0   & 0  & $5.6\times10^{-17}$ & $4.8\times10^{0}$ & $0$ & 0 \tabularnewline
				\hline
				
			\end{tabular}
		\end{table*}
		
		In addition to the displacement of the $E_\mathrm{rf,r}$ minimum and residual RF field along the trap axis, apertures in the trap electrodes also modify the curvature of the RF pseudopotential and thereby the secular frequencies. Unlike micromotion, these secular-frequency changes cannot be compensated locally without modifying the trapping potential. For unfavorable aperture placements, relative secular-frequency changes of up to about 5\% already occur for an aperture width of $w_\mathrm{a}=30~\mu\mathrm{m}$. For larger apertures, this effect becomes substantially stronger, reaching changes of up to approximately 20\% for $w_\mathrm{a}=100~\mu\mathrm{m}$. These aperture-induced modifications of secular frequency therefore constitute an additional design constraint for photonic-integrated surface traps.

		With a fixed aperture width, increasing the thickness of the electrode from 1 $\mu$m to 20 $\mu$m can reduce field distortions by two orders of magnitude, due to the reduction of the edge effect. However, care must be taken to ensure that the electrode does not obstruct the outcoupled laser beam, unless a technique is developed to shape the metal layer of the electrode at the angle of the outcoupled beam.

		Arranging apertures symmetrically provides an effective strategy for compensating the displacement of the $E_\mathrm{rf,r}$ minimum and the phase difference between the RF electrodes. Nevertheless, this approach introduces stronger residual RF fields along the trap axis, if additional apertures are required or if the aperture size must be increased to maintain symmetry. Independent of the presence of an aperture in an electrode, it is not possible to fully compensate the out-of-plane component (x) of the RF field at the target ion position due to the inherently two-dimensional geometry of planar surface traps. 
		
		The application of TCO coatings on apertures significantly reduces the potential and phase distortions on the electrodes, thereby strongly reducing  RF field distortions at the target ion position, e.g. from $E_{\mathrm{rf,y}}> 900$ V/m to $E_{\mathrm{rf,y}}< 100$ V/m. Only small residual RF field components remain due to the topography of the modified surface. Future technological developments could focus on leveling the topography of surface traps to reach a flat surface with TCO coating.

		Although the present study is limited to simulated environments, the outcomes provide a basis for anticipating challenges in real-world applications. We explore the implications of the simulation results for precision measurements and trapped-ion quantum technologies. For trapped ion based optical clocks, the time dilation shift due to excess micromotion is one common contribution to the uncertainty. The fractional time dilation shift due to EMM is given by~\cite{Kell2019}
		\begin{equation}
			\left| \frac{\Delta \nu}{\nu_0} \right|_{\mathrm{TD}}
			= \left( \frac{e E_{\mathrm{rf}}}{2 m c \Omega_{\mathrm{rf}}} \right)^2 ,
			\label{eq:micromotion}
		\end{equation}
		where $e$ is the elementary charge, $m$ is the ion mass and $c$ is the speed of light.

		Besides excess micromotion, the field distortion caused by an aperture also leads to ion heating through RF noise. The RF noise induced heating rate can be estimated using~\cite{Blakestad2009, Kali2021}
		\begin{equation}
			\dot{\bar n}_i
			=
			\frac{e^{4}}{16\,m^{3}\Omega_{\mathrm{rf}}^{4}\hbar\,\omega_{\mathrm{sec},i}}
			\left(
			\frac{\partial}{\partial i}
			E_{\mathrm{rf}}^{2}(\mathbf r)
			\right)^{2}
			\frac{S_V(\Omega_{\mathrm{rf}}\!\pm\!\omega_{\mathrm{sec},i})}{U_\mathrm{RF}^{2}},
			\label{eq:heating_component}
		\end{equation}
		where $S_V$ is the RF-voltage noise spectral density at $\Omega_{\mathrm{rf}}\ \pm\ \omega_{\mathrm{sec},i}$ and $\hbar$ the reduced Planck constant.  
		
		\begin{table*}[t]
			\centering
			
			\setlength{\tabcolsep}{3pt}
			
			\caption{Electric field components, micromotion-induced fractional time-dilation shift 
				$\left| \Delta\nu/\nu_0 \right|_{\mathrm{TD}}$, and RF-noise heating rates along the three motional
				directions, evaluated at the outermost ion located at $z = 17.5~\mu\mathrm{m}$ in a crystal of ten
				$^{172}\mathrm{Yb}^+$ ions. 
				All calculations assume radial and axial secular frequencies of 
				$\nu_{\mathrm{sec}, r} = 1.9\,\mathrm{MHz}$ and $\nu_{\mathrm{sec},z} = 300\,\mathrm{kHz}$, respectively.
				The position of the single aperture in the RF electrode is fixed at 
				$p_y = 126.8~\mu\mathrm{m}$ and $p_z = 0~\mu\mathrm{m}$. 
				Four configurations are considered: 
				(i) a single-aperture geometry; 
				(ii) a single aperture with an ITO coating; 
				(iii) a symmetric two-aperture geometry obtained by introducing a second aperture 
				mirror-symmetric with respect to the $z$ axis; and 
				(iv) the symmetric configuration with ITO coating.}
			
			\label{tab:dup_blocks_z=17p5}
			
			\begin{tabular}{|p{2.0cm}|c|c|c|c|c|c|c|c|}
				\hline
				\multicolumn{2}{|c|}{} 
				& \multicolumn{7}{c|}{Evaluated at the target ion position} \\
				\hline
				\centering Configuration
				& \shortstack{$w_a$ \\ ($\mu$m)}
				& \shortstack{$E_x$ \\ (V/m)}
				& \shortstack{$E_y$ \\ (V/m)}
				& \shortstack{$E_z$ \\ (V/m)}
				& $\left| \Delta\nu/\nu_0 \right|_{\mathrm{TD}}$
				& \shortstack{$\dot{\bar n}_x$ \\ (s$^{-1}$)}
				& \shortstack{$\dot{\bar n}_y$ \\ (s$^{-1}$)}
				& \shortstack{$\dot{\bar n}_z$ \\ (s$^{-1}$)} \tabularnewline
				\hline
				
				\multirow{3}{*}{\centering single}
				& 30   & -90.0   & 963.7   & -131.8   & $8.3\times10^{-17}$ 
				& $5.4\times10^{-2}$ & $5.9\times10^{0}$ & $9.3\times10^{-4}$ \tabularnewline
				& 77.5 & -2213.7 & 15363.4 & -2106.9  & $2.1\times10^{-14}$ 
				& $5.4\times10^{1}$ & $1.3\times10^{3}$ & $5.3\times10^{1}$ \tabularnewline
				& 100  & -5412.8 & 29916.9 & -4061.3  & $8.1\times10^{-14}$ 
				& $4.4\times10^{2}$ & $4.1\times10^{3}$ & $7.2\times10^{2}$ \tabularnewline
				\hline
				
				\multirow{3}{*}{\centering \shortstack{single \\ + ITO}}
				& 30   & -0.9   & 85.9   & -13.7   & $6.6\times10^{-19}$ 
				& $2.1\times10^{-5}$ & $4.9\times10^{-2}$ & $1.1\times10^{-7}$ \tabularnewline
				& 77.5 & -142.0 & 950.7  & -134.5  & $8.2\times10^{-17}$ 
				& $1.3\times10^{-1}$ & $5.8\times10^{0}$ & $8.0\times10^{-4}$ \tabularnewline
				& 100  & -374.8 & 1911.0 & -251.6  & $3.3\times10^{-16}$ 
				& $9.5\times10^{-1}$ & $2.3\times10^{1}$ & $1.2\times10^{-2}$ \tabularnewline
				\hline
				
				\multirow{3}{*}{\centering symmetry}
				& 30   & -167.7   & 0 & -281.8    & $9.4\times10^{-18}$ 
				& $1.7\times10^{-1}$ & $0$ & $1.8\times10^{-3}$ \tabularnewline
				& 77.5 & -4391.0  & 0 & -4225.2   & $3.2\times10^{-15}$ 
				& $1.0\times10^{2}$ & $0$ & $1.0\times10^{2}$ \tabularnewline
				& 100  & -10784.5 & 0 & -8150.3  & $1.6\times10^{-14}$ 
				& $5.8\times10^{2}$ & $0$ & $1.2\times10^{3}$ \tabularnewline
				\hline
				
				\multirow{3}{*}{\centering \shortstack{symmetry \\ + ITO}}
				& 30   & 9.0    & 0 & -42.4   & $2.0\times10^{-19}$ 
				& $7.6\times10^{-4}$ & $0$ & $2.8\times10^{-7}$ \tabularnewline
				& 77.5 & -258.0 & 0 & -267.0  & $1.2\times10^{-17}$ 
				& $4.1\times10^{-1}$ & $0$ & $1.3\times10^{-3}$ \tabularnewline
				& 100  & -749.7 & 0 & -528.0  & $7.3\times10^{-17}$ 
				& $4.0\times10^{0}$ & $0$ & $1.4\times10^{-2}$ \tabularnewline
				\hline
				
			\end{tabular}
		\end{table*}

		As shown throughout this work, the aperture displaces the RF-field minimum
		from the geometric trap center. For single ion, the residual RF field can be
		compensated by applying DC voltages to shift the ion to the RF-field
		minimum. In this case, the integrated outcoupling optics must be foreseen and designed
		to accommodate this displacement, ensuring alignment between the ion
		position and the optical beam.
		Table~\ref{tab:dup_blocks_z=0} summarizes the corresponding RF-minimum
		displacements for aperture widths of $w_a = 30~\mu\mathrm{m}$,
		$77.5~\mu\mathrm{m}$, and $100~\mu\mathrm{m}$. For each aperture width,
		four configurations are considered: a single aperture and a symmetric
		two-aperture arrangement obtained by mirroring the aperture  with respect to the z-axis, each with and without an ITO coating.
		
		If the micromotion can  not be compensated by shifting the ion to the RF-field minimum, the residual
		RF field results in a time-dilation shift and RF-noise--induced heating, which are also listed in Table~\ref{tab:dup_blocks_z=0}. The calculation is performed for a single $^{172}\mathrm{Yb}^+$ ion,
		assuming a radial secular frequency of
		$\nu_{\mathrm{sec}, r} = 1.9\,\mathrm{MHz}$ and an axial secular frequency
		of $\nu_{\mathrm{sec},z} = 300\,\mathrm{kHz}$. 
		The RF-noise–induced heating rate is estimated following the model and experimental observations reported by
		Kalincev \textit{et al.}~\cite{Kali2021}. 
		We assume a resonant RF circuit with quality factor $Q = 542$ and
		inductance $L = 2.5\,\mu\mathrm{H}$, and a constant power spectral
		density of the input RF noise $S_P = 1.5\times10^{-14}\,\mathrm{W\,Hz^{-1}}$.
		The resulting RF-voltage noise spectral densities  are
		$1.23\times10^{-13}\,\mathrm{V^2\,Hz^{-1}}$ and
		$4.93\times10^{-12}\,\mathrm{V^2\,Hz^{-1}}$ for the radial and axial
		modes, respectively.

		For
		single-ion addressing with a tightly focused beam, larger apertures
		without ITO coating lead to larger displacements, which can cause a
		significant mismatch between the ion position and the optical beam waist
		if not properly considered during the design phase. Symmetric
		aperture configurations do not fully cancel the residual RF field.
		
		So far, we have considered single-ion addressing at the trap minimum.
		We now estimate the micromotion-induced fractional time-dilation shift
		$\left| \Delta\nu/\nu_0 \right|_{\mathrm{TD}}$ and the RF-noise--induced
		heating rates for a spatially extended linear ion crystal along the trap axis.
		Assuming the same secular frequencies as used in the single-ion calculation, we consider a crystal of ten $^{172}\mathrm{Yb}^+$ ions with an axial length of approximately $35~\mu\mathrm{m}$. The equilibrium ion positions are calculated following the method of James~\cite{james1997quantum}. As a representative example, we take the outermost ion at $z=17.5~\mu\mathrm{m}$ and evaluate the residual RF electric field, the micromotion-induced fractional time-dilation shift, and the RF-noise heating rate at this position. The corresponding results are
		summarized in Table~\ref{tab:dup_blocks_z=17p5}.
		
		For symmetric aperture configurations, the micromotion-induced
		time-dilation shift is not fully cancelled, but is reduced compared to the asymmetric case. Regarding heating, the
		symmetric configuration cancels the RF-noise--induced heating in the
		$y$ direction, but increases the heating rates in the $x$ and $z$
		directions.
		
		Applying an ITO coating to the aperture significantly reduces the field
		distortion and thereby decreases the fractional time-dilation shift by
		approximately two orders of magnitude. The corresponding RF-noise--induced
		heating rates are reduced by two to four orders of magnitude, depending
		on the motional direction. This demonstrates that both the time-dilation
		shift and the RF-noise--induced heating arising from aperture-induced
		field distortions can be effectively mitigated by covering the apertures
		with ITO.\\

		\begin{acknowledgments}
			The authors thank Karan Mehta for the valuable discussions about the trap chip design, Carl-Frederik Grimpe and Fatemeh Salahshoori for their insights on the design constraints imposed by grating couplers and fabrication tolerances for nanophotonics, and Markus Kromrey for constructive discussions on the FEM simulations. GD acknowledges the Ministry of Science and Culture of Lower Saxony (MWK) and the Volkswagen foundation for funding the QVLS-Q1 project No. 51171233. This work was supported by the BMBF within the collaborative project ATIQ Grant No. 13N16126, and the Deutsche Forschungsgemeinschaft (DFG, German Research Foundation) under Germany’s Excellence Strategy – EXC-2123 QuantumFrontiers – 390837967.\\
		\end{acknowledgments}

	\end{document}